\documentclass[reprint,aps,prd,longbibliography]{revtex4-1}
\usepackage{graphicx}
\usepackage{dcolumn}
\usepackage{bm}
\usepackage{amssymb}
\usepackage{multirow}
\usepackage{color}
\usepackage{hyperref}
\usepackage{ulem}
\usepackage{float}
\usepackage{amsmath}
\usepackage{appendix}

\begin{document}

\title{Self-assembly Monte Carlo reveals localized entanglement in giant polymer melts}

\author{Enrico Fornasa$^1$}
\affiliation{(1) Scuola Internazionale Superiore di Studi Avanzati (SISSA), Via Bonomea 265, 34136 Trieste, Italy}

\author{Francesco Slongo$^1$}
\affiliation{(1) Scuola Internazionale Superiore di Studi Avanzati (SISSA), Via Bonomea 265, 34136 Trieste, Italy}

\author{Cristian Micheletti$^1$}
\email{Corresponding author: cristian.micheletti@sissa.it}
\affiliation{(1) Scuola Internazionale Superiore di Studi Avanzati (SISSA), Via Bonomea 265, 34136 Trieste, Italy}

\begin{abstract}
Topological entanglements are central to understanding and predicting the properties of polymer melts. Yet, they make equilibrium sampling computationally challenging, as decorrelation times grow rapidly with chain length. Here, we introduce a Monte Carlo scheme that bypasses typical computational bottlenecks by working in a self-assembly ensemble rather than at fixed composition. Strictly local moves efficiently propagate backbone reconnections across scales while conserving the number of linear chains, achieving near-linear scaling of decorrelation time with system size, $\tau_{\rm eq}\sim V^{1.0}$. With this method, formulated for a fully-packed lattice, we equilibrate periodic systems totalling up to $\simeq 1.1\times 10^9$ monomers, accessing a universal melt regime insensitive to lattice details. We analyze intra- and inter-chain entanglements for chains of up to $N\simeq 5 \times 10^5$ monomers, revealing that they manifest as localized knots and links rather than as global tangles. Finally, we show that the magnitude of the Gauss linking integral between neighbouring chains grows only as $N^{1/4}$.
\end{abstract}

\maketitle

\noindent

{\bf Introduction.}
Polymer melts owe many of their physical properties to topological entanglements established by the densely packed chains. Examples range from the structural plasticity of eukaryotic genomes to the viscoelasticity of synthetic materials and the mechanics of interlocked filaments~\cite{grosberg1988role,everaers2017flory,doi:10.1021/acs.macromol.8b01836,orlandini2021topological,wang2023entanglements,slongo2023quantum,doi:10.1021/acspolymersau.2c00049,tubiana2024topology,grosse2025scale}. The regime most relevant to soft matter, materials science, and biological physics occurs when the chain length, $N$, largely exceeds the entanglement length, the typical contour separation between entanglement points~\cite{everaers2004rheology,hoy2020unified}. In this regime, equilibration is governed by the slow reptation of the individual filaments within the constraining mesh, with disengagement times growing at least as $O(N^3)$~\cite{doi1988theory,rubinstein2003polymer,mcleish2002tube,abadi2018entangled}.
In molecular dynamics simulations with cell or neighbour lists, the additional cost of computing interactions leads to wall-clock decorrelation times scaling as $\gtrsim O(N^4)$~\cite{dunweg1998molecular}. Analogous considerations hold for Monte Carlo methods with topology-preserving local moves. Consequently, characterizing and predicting the properties of dense polymer melts with realistic simulations become impractical as $N$ grows well beyond the entanglement length.

To make the problem more tractable, several accelerated sampling strategies have been proposed, including hierarchical coarse-graining, thermodynamic reweighting, and enrichment~\cite{rosenbluth1955monte,tesi1996monte,deutsch1997long,grassberger1997pruned,prellberg2004flat,lua2004fractal,lua2005under,siepmann1992configurational,clisby2010accurate,micheletti2011polymers,ghosh2018epigenome,rubinstein1987discretized,ma8018624,hugouvieux2009amphiphilic,schram2018simulation,karayiannis2002novel,reith2010implementation,dietz2022facile,bonato2022topological,svaneborg2023multiscale}. Off-lattice bridging moves that cut and rejoin bonds while conserving melt composition (i.e.~polymer lengths, chain count and density), can bypass topological constraints and reduce the effective dynamic exponents to the 2-2.5 range~\cite{mavrantzas1999end,auhl2003equilibration,svaneborg2016multiscale}. On lattices, local reconnection moves for space-filling walks can achieve $O(N^2)$ decorrelation scaling~\cite{deutsch1997long,mansfield2006unbiased}, and efficient hierarchical construction algorithms can be employed too~\cite{smrek2013novel,halverson2014melt}. Despite these advances,
most approaches to equilibrating melts are formulated for the fixed-composition ensemble, where enforcing constraints of chain lengths and connectivity remains a major computational bottleneck.

Here, we introduce an efficient self-assembly Monte Carlo (SAMC) scheme for dense polymers on a cubic lattice at full packing. The system, composed of a periodic ring-linear blend, is evolved with local endpoint bond-swaps that allow the composition to fluctuate while still conserving density, self-avoidance, and the number $n$ of linear chains. By allowing reconnections, SAMC avoids the aforementioned expensive backbone tracing checks that hinder sampling methods at fixed-composition. As a result, the global decorrelation time, measured in Monte Carlo steps, scales approximately linearly with system size, $\tau_{\rm eq} \sim V^{1.001 \pm 0.002}$. Because the CPU time per Monte Carlo step varies only weakly with $V$, the near-linear scaling applies also to wall-clock decorrelation time.

This upends the usual hierarchy of computational bottlenecks: while for conventional methods the most onerous step is the collection of uncorrelated states, for SAMC it is instead their analysis, as the computational cost of calculating physical observables often scales quadratically or worse  with chain lengths.

In the stationary SAMC ensemble, giant linear components spontaneously emerge against a background of finite and non-extensive rings. In fact, the $n$ linear chains are individually macroscopic, with average lengths $\langle N \rangle$ proportional to the system volume, which is almost entirely ($\sim 90\%$) occupied by them.

\begin{figure*}[t!]
\centering
\includegraphics[width=\textwidth]{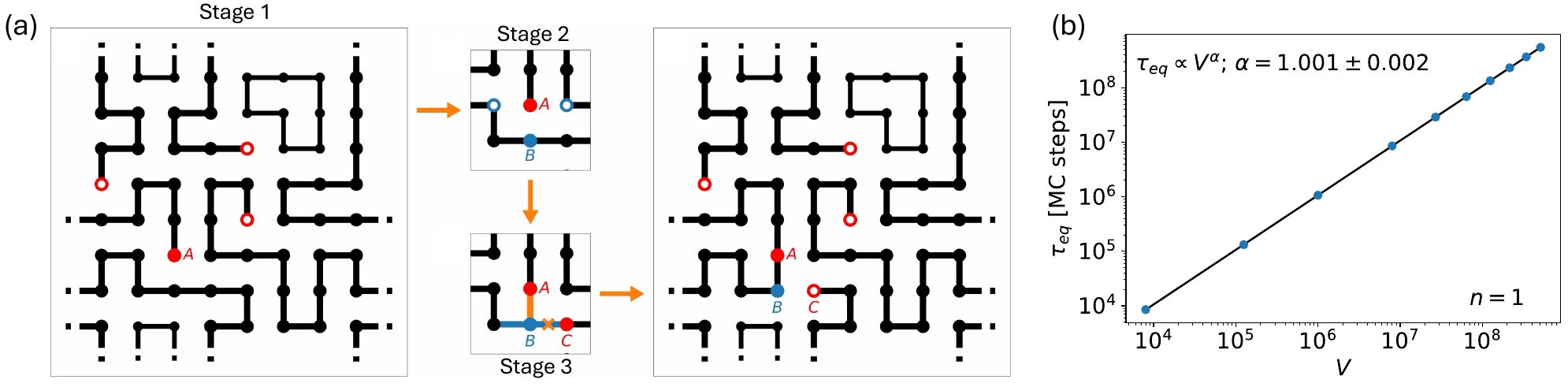}
\caption{{\bf Endpoint bond-swap move and decorrelation time.} (a) Sketch of the three stages involved in an endpoint bond-swap move. For clarity, the sketch is shown for a blend with $n=2$ linear chains on a two-dimensional ($8\times 8$) periodic square lattice. First, an endpoint $A$ is randomly chosen among the $2n = 4$ degree-1 sites (open circles). Next, a site $B$ is randomly chosen among the non-bonded nearest neighbours of $A$. Then, a site $C$ is randomly chosen among the bonded neighbours of $B$. If $C$ is an endpoint, the move is rejected, otherwise bond $BC$ is replaced by $AB$, obtaining the evolved configuration shown on the right. The move conserves packing, self-avoidance, and the total number of linear chains, while chain lengths and ring counts typically vary. (b) Global decorrelation time $\tau_{\rm eq}$ for systems with a single linear chain ($n=1$) per periodic box as a function of the box volume $V= L^3$. $\tau_{\rm eq}$ was obtained as the integrated autocorrelation of edge-state overlap $q$, and is measured in Monte Carlo steps. The power law best fit yields a practically linear dependence of $\tau_{\rm eq}$ on $V$, as indicated, with the prefactor equal to $1.05 \pm 0.01$. The relative statistical uncertainty on $\tau_{\rm eq}$ is estimated to be smaller than 1\% (Supplementary Figure 3 and Supplementary Note 5).}
\label{fig:fig1_model}
\end{figure*}

Leveraging the efficiency of SAMC and the emergence of giant components, we equilibrate fully packed systems in which only a few giant chains collectively comprise up to $\simeq 1.1\times10^9$ monomers. By analyzing chains with millions of monomers, we show that their metric statistics are consistent with those of pure linear melts, indicating that their scaling properties lie in a universal melt regime, insensitive to microscopic lattice details.

We thus push the topological characterization of melts to previously unexplored chain densities and lengths. We profile intra- and inter-chain entanglements across scales and show that they manifest as localized knots and links rather than as global tangles. Finally, we show that the characteristic Gauss linking integral between neighbouring chains grows only as $N^{1/4}$.

\section*{Results}
{\bf Self-assembly Monte Carlo.} We study self-assembled blends of linear chains and rings that are fully packed on a simple-cubic lattice. The system is evolved by a Monte Carlo scheme with strictly local endpoint bond swaps that efficiently propagate backbone reconnections across scales while conserving the number of linear chains. This self-assembly Monte Carlo (SAMC) method has near-linear global decorrelation time with system size and yields macroscopic linear chains with melt-like statistics. This makes it possible to efficiently  equilibrate very large systems and thus investigate and establish large-scale properties that are insensitive to microscopic lattice details.

We consider a fully-packed ring-linear blend on a $L\times L \times L$ periodic lattice, containing $n$ linear chains. Accordingly, all sites have connectivity degree $2$ except those hosting the $2n$ endpoints of the chains, which have degree 1.

\begin{figure*}[t!]
\centering
\includegraphics[width=0.95\textwidth]{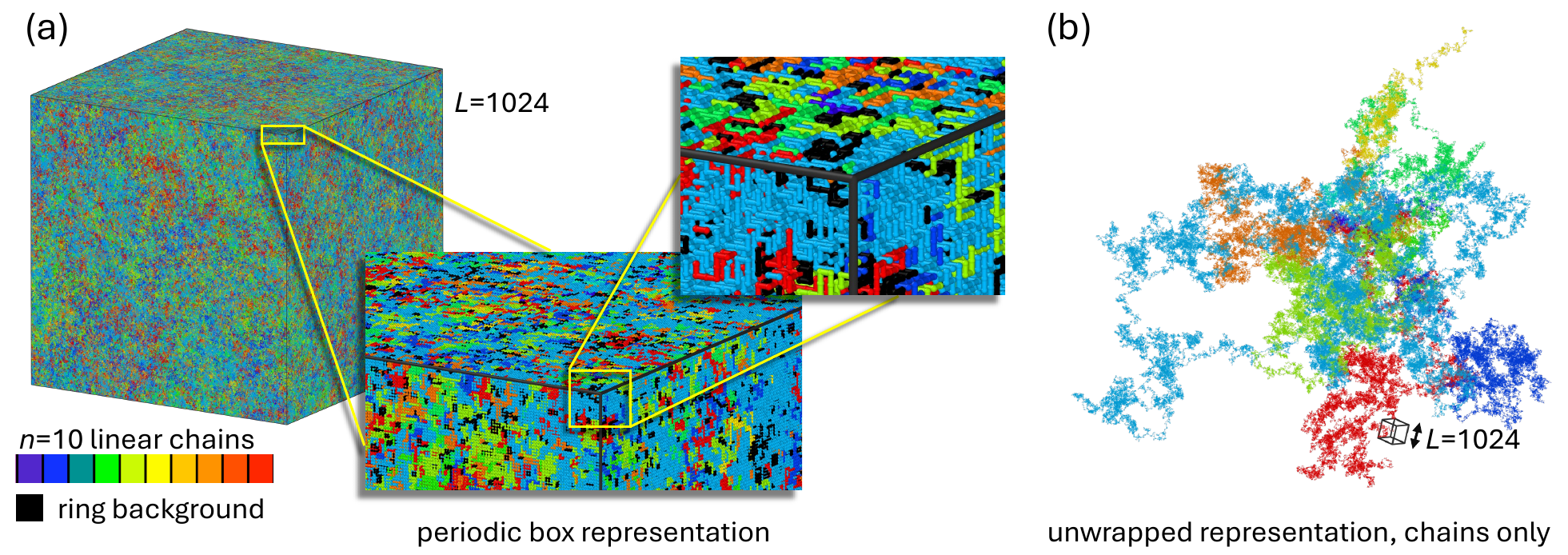}
\caption{{\bf Emergent giant linear components.} (a) Wrapped and (b) unwrapped representations of an SAMC-equilibrated configuration with $n=10$ linear chains on a fully-packed $1024\times1024\times1024$ periodic lattice, totalling $\simeq 1.1 \times 10^9$ monomers. The $10$ linear chains are distinguished by a rainbow palette, while the ring background is shown in black. Collectively, the linear chains comprise $\simeq 9.7 \times 10^8$ monomers, about 90\% of the volume.}
\label{fig:fig2_config}
\end{figure*}
The elementary SAMC move is sketched in Figure \ref{fig:fig1_model}a and is articulated over the following stages. First, one of the $2n$ endpoints, denoted $A$, is chosen uniformly. Next, a site $B$ is chosen uniformly among the non-bonded nearest neighbours of $A$. Then, a site $C$ is chosen uniformly among the bonded neighbours of $B$ (one choice if $B$ is an endpoint, two otherwise). If $C$ is an endpoint, the move is rejected and the current configuration is retained, otherwise bond $BC$ is replaced with $AB$.

After each accepted move, every site retains its degree except for sites $A$, whose degree changes from 1 to 2 after the addition of bond $AB$, and site $C$, whose degree changes from 2 to 1 after removing bond $BC$. As a result, $C$ replaces $A$ as the free end. Rejecting proposals in which $C$ is an endpoint is necessary because the bond swap would disconnect that site, violating the conservation of the number of free ends. With the above move set, such rejections occur with probability $O(n/V)$, where $V=L^3$ is the lattice volume, and hence become negligible as $V$ grows at fixed $n$ (Supplementary Figure 2). An optional lower-rejection variant, discussed in Supplementary Note 6, reduces this scaling to $O(n^2/V^2)$.

Thus, the SAMC evolution conserves full packing, self-avoidance, and the number of endpoints (and thus the number of linear chains, $n$). Instead, the linear chains' lengths, and the number and lengths of rings can fluctuate.

The move proposal is symmetric on admissible endpoint bond swaps, and all such proposals are accepted. The rejected proposals instead contribute only to self-loops. Therefore the resulting Markov chain satisfies detailed balance with respect to a uniform stationary measure, see Supplementary Note 1. On the considered periodic hypercubic lattices at full packing, the dynamics is free of blocked (immobile) configurations except in the limiting case $n=V/2$, where the system is made entirely of chains of length 1 (Supplementary Note 2). While a general proof of ergodicity is lacking, comparisons against exhaustively enumerable systems with up to $\simeq 10^5$ states show that a single extensive SAMC run samples uniformly the full configuration space (Supplementary Figure 1). These results support the effective ergodicity and uniformity of SAMC.

{\bf Global decorrelation: near-linear scaling.}
For a conservative measure of global decorrelation, we take the integrated autocorrelation time of the edge overlap, $\tau_{\rm eq}$, expressed in Monte Carlo steps. This overlap measures the fraction of lattice edges that have the same state (empty or occupied) in two configurations (Methods and Supplementary Note 5).

To characterize the scaling of $\tau_{\rm eq}$ with system size, we computed $\tau_{\rm eq}$ for $L$ ranging from $20$ to $800$, corresponding to $V$ spanning more than four decades, from $8.0\times10^3$ to $5.12\times10^8$. A power-law fit of the data yields a practically linear scaling of the decorrelation time with system size, $\tau_{\rm eq} \propto V^{1.001 \pm 0.002}$ (Figure \ref{fig:fig1_model}b). The near-linear decorrelation reflects three features of SAMC. First, working in a self-assembly ensemble circumvents constraints inherent to the conventional fixed-composition ensemble. Second, the endpoint bond-swap move is local and yet it efficiently propagates connectivity changes across scales. Lastly, the SAMC evolution requires tracking only the $2n$ endpoints, dispensing with the onerous tracing of entire backbones that fixed-composition accelerated schemes require at each bond swap. With SAMC, backbone tracing is needed only at the analysis stage, and exclusively for the independent samples used to compute observables of interest. In this regard, we collect configurations at time intervals of $3\tau_{\rm eq}$ to ensure their statistical independence.

In terms of wall-clock runtime, the single-core CPU time per Monte Carlo step on standard workstations is approximately independent of $V$ and equal to $\simeq 0.2$ $\mu$s as long as RAM capacity is not exceeded (Supplementary Figure 5). Accordingly, the near-linear scaling of $\tau_{\rm eq}$ on $V$ is carried over to the wall-clock decorrelation time. On the tested workstations, decorrelating configurations of $100\times100\times100$ systems ($V=10^6$) at $3\tau_{\rm eq}$ intervals requires $\simeq 0.6$ s.

Finally, regarding the analysis, we note that common observables of interest, including those used later to study entanglement, have an $O(N^2)$ cost or worse. As a consequence, for the large systems considered here, the dominant computational bottleneck shifts from equilibration to post-processing analysis.

{\bf Giant linear components.} An emergent property of the SAMC ensemble is that the $n$ linear chains are individually giant or macroscopic, i.e.~with length growing proportionally to $V$. By contrast, the complementary ring background consists of numerous finite loops.

Figure \ref{fig:fig2_config} shows a typical SAMC-equilibrated configuration of a ring-linear blend with $n=10$ chains fully occupying a $1024 \times 1024 \times 1024$ periodic lattice ($V\simeq 1.1\times10^9$), the largest addressable system with the RAM available to us. We note that this size exceeds the maximum one used for the decorrelation analysis of Figure \ref{fig:fig1_model}b because the latter required the simultaneous storage of multiple configurations in memory, while SAMC sampling requires storing only one configuration. Figure \ref{fig:fig2_config}a shows the blend with coordinates wrapped around the boundaries of the periodic $L\times L\times L$ box. The linear chains are manifestly the giant components of the system, collectively occupying $\simeq 9.7 \times 10^8$ monomers, or 90\% of the box volume. Figure \ref{fig:fig2_config}b shows one of the possible unwrapped representations of the same linear chains in panel a, obtained by selecting a reference periodic box and then tracing each backbone continuously across periodic boundaries, starting from one of its ends in that box.

\begin{figure*}[t!]
\centering
\includegraphics[width=\textwidth]{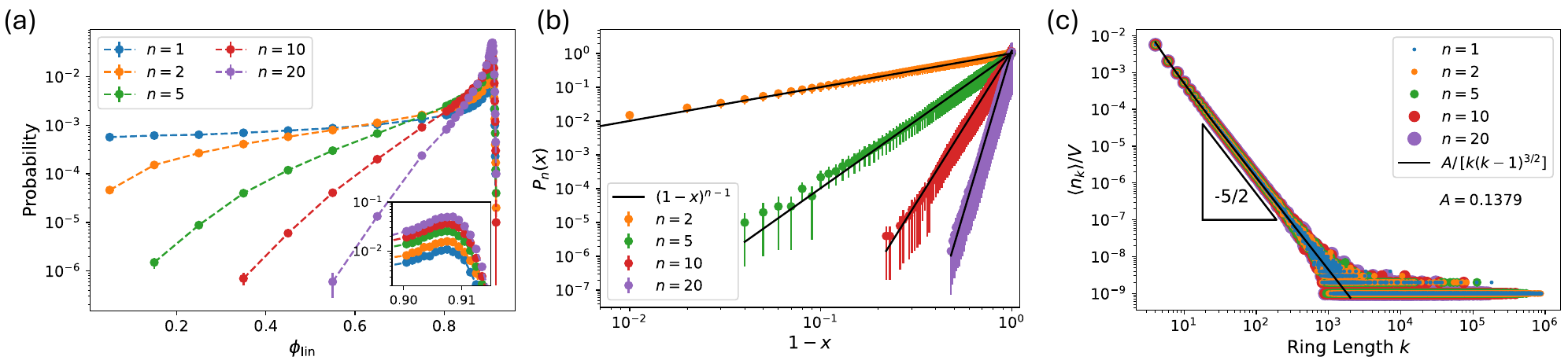}
\caption{{\bf Statistics of the macroscopic linear chains and ring background.}
 (a) Probability distribution of $\phi_{\rm lin}$, the volume fraction collectively occupied by the indicated number of linear chains, $n$. Inset: close-up near the modal value, $\bar{\phi}_{\rm lin} \simeq 0.9$. (b) Tail probability, $P_n(x)$, that individual linear chains at $\bar{\phi}_{\rm lin}$ occupy a portion larger than $x$ of the linear volume, $V_{\rm lin}=\bar{\phi}_{\rm lin} V$. The $P_n(x)$ data are well approximated by the theoretical Dirichlet distribution (solid lines) for random $n$-partitioning of the linear volume, $P_n^{\rm theor}(x)=(1-x)^{n-1}$. (c) Average number density of rings of length $k$, $\langle n_k \rangle/V$, for configurations conditioned at $\bar{\phi}_{\rm lin}$. The solid line is the parameter-free theoretical prediction $\langle n_k \rangle^{\rm theor}/V= A/[k(k-1)^{3/2}]$, with $A \simeq 0.1379$ (Supplementary Note 10). All plots are for independent SAMC-equilibrated samples of $100\times100\times100$ systems. For each $n$, $10^5$ independent configurations were used for panel (a) and $10^3$ independent conditioned configurations for panels (b,c). Error bars in panels (a) and (b) represent standard errors.}
\label{fig:fig3_giants}
\end{figure*}

Indeed, for the considered range of $n$, the distribution of the volume fraction occupied by the linear chains $\phi_{\rm lin}$ converges with increasing $V$ to an asymptotic curve sharply peaked at $\bar{\phi}_{\rm lin}\simeq 0.9$ (Supplementary Figure 7 and Supplementary Table 1). The $\phi_{\rm lin}$ distributions for the effectively asymptotic $100 \times 100 \times 100$ system are shown in Figure \ref{fig:fig3_giants}a. The tails beyond the peak have a common rapid decay while the shoulders at $\phi_{\rm lin} < \bar{\phi}_{\rm lin}$ become visibly sharper with increasing $n$. Thus, the combined linear components are most likely to occupy about 90\% of the lattice sites across the explored $n$.

Analysis of how the collectively macroscopic linear volume is partitioned among the $n$ linear chains shows that the latter are individually macroscopic. To establish this, we first conditioned the system to the macroscopic linear regime ($\bar{\phi}_{\rm lin}$) by retaining from the independent configurations only those with $\phi_{\rm lin}\in[0.90,0.91]$. From the individual chain lengths, $N_{i=1,...,n}$, we next computed the corresponding linear volume portions, $x_i=N_i/V_{\rm lin}$, where $V_{\rm lin}=\bar{\phi}_{\rm lin} V$.

For $n=1$, the conditioning clearly yields a near monodisperse chain length, $N_1\simeq 0.9\,V$, hence macroscopic. To characterize the $n>1$ case, we consider the tail probability $P_n(x)$ that $x_i$ is larger than $x$ and compare it with the Dirichlet expression for uniform random $n$-partitioning of the unit interval, $P_n^{\rm theor}(x)=(1-x)^{\,n-1}$. Figure \ref{fig:fig3_giants}b shows that the observed $P_n(x)$ follows closely the theoretical expression. Notice that the agreement is not the result of a best fit procedure, since the theoretical expression has no adjustable parameters.

We conclude that, at the modal linear volume fraction $\bar{\phi}_{\rm lin}$, the $n$ linear chains are statistically equivalent and individually giant or macroscopic, with average lengths scaling as $O(V/n)$, see also Supplementary Figure 8.

By contrast, the ring background conditioned at $\bar{\phi}_{\rm lin}$ is instead finite and dominated by short rings. Figure \ref{fig:fig3_giants}c shows that the average number density $\langle n_k \rangle /V$ of rings of length $k$ collapses on the solid curve, corresponding to the theoretical prediction $\langle n_k \rangle^{\rm theor}/V= A/[k(k-1)^{3/2}]$, with $A \simeq 0.1379$. This expression is based on the statistics of loops formed by chain closure in melt-like conditions and, notably, has no free parameters, Supplementary Note 10. Consistent with the theoretical scaling, the data for $k \gg 1$ decay rapidly as $\langle n_k \rangle/V \sim k^{-5/2}$. As a consequence, the average ring length at $\bar{\phi}_{\rm lin}$ remains finite across all considered $n$ and $V$ and stays close to the theoretical estimate $\langle l_{\rm ring}^{\rm theor} \rangle \simeq 8.26$ (Supplementary Note 10 and Supplementary Table 3).

Two observations are in order. First, the regime of giant chains must eventually break down at sufficiently large $n$, since for $n=O(V)$ the typical chain length is necessarily $O(1)$. Second, our observation of a giant linear component at $n=1$ connects to recent mathematical developments for the three-dimensional double-dimer model, for which rigorous proofs have established the existence of a macroscopic open path upon insertion of a single pair of defects~\cite{quitmann2023macroscopic}. While the double-dimer model and ours differ for both the configurational averaging and the nature of the ring background, our results provide a direct physical realization of this prediction, and offer a quantitative basis for motivating extensions of the rigorous proof to $n>1$.

\begin{figure*}[t!]
\centering
\includegraphics[width=0.8\textwidth]{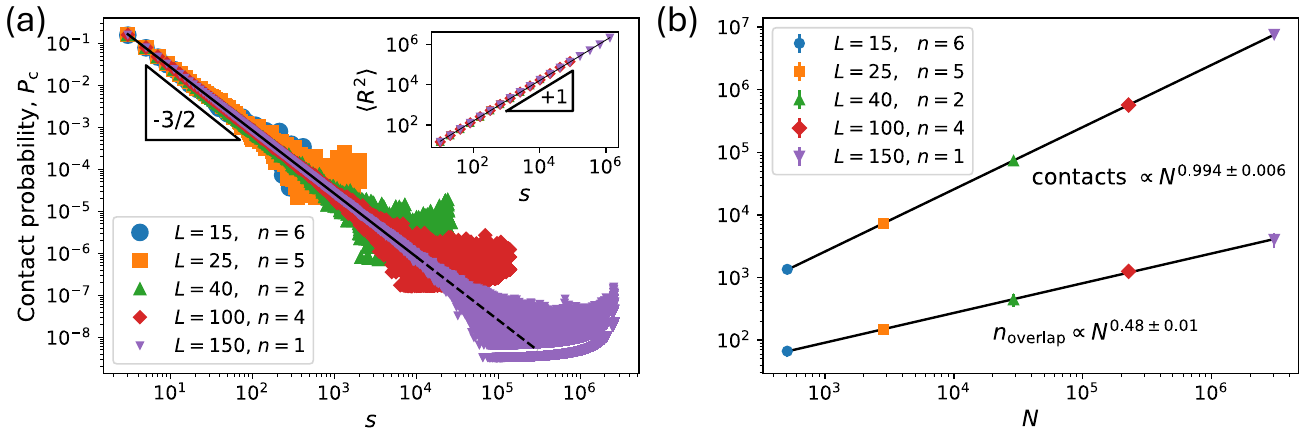}
\caption{{\bf Melt-like metric scaling of giant linear chains.} (a) Contact probability $P_c$ and mean square internal distance $\langle R^2\rangle$ (inset) of giant linear chains as a function of contour separation $s$. Power-law best fits (solid lines) to $P_c$ and $\langle R^2 \rangle$ data ($L=150$, $n=1$, $s < 10^4$) yield respectively the exponents $-1.505 \pm 0.003$ and $1.00 \pm 0.01$, consistent with melt behaviour. (b) Average total number of inter-chain monomeric contacts and average number of distinct overlapping chains as a function of chain length, $N$. Again, the indicated exponents of power-law best fits (solid lines) are consistent with melt behaviour. Data are for various system sizes and numbers of linear chains $n$; additional combinations are in Supplementary Figure 9. For each combination of size and $n$, data are from $10^2$ independent configurations conditioned at $\bar{\phi}_{\rm lin}$ and with near-monodisperse lengths, $N \simeq \bar{\phi}_{\rm lin} V/n$. Error bars in panel (b) represent standard errors.}
\label{fig:fig4_metric}
\end{figure*}

{\bf Giant linear melts.}
Consistent with the fact that the linear chains are the sole macroscopic components at $\bar{\phi}_{\rm lin}$, we observe that their metric properties follow closely the scaling laws expected for melts over the investigated range.

Figure \ref{fig:fig4_metric}a shows the intra-chain contact statistics at $\bar{\phi}_{\rm lin}$ for selected values of $n$ and $V$; additional combinations are in Supplementary Figure 9. The data are for the probability, $P_c(s)$, that pairs of monomers at contour distance $s$ are in contact, i.e.~on nearest neighbouring sites. To connect with known scaling properties for melts, the analysis was performed on near monodisperse configurations. Specifically, the configurations were selected a posteriori such that each of the $n$ chain lengths fell within 10\% of the equisubdivision value, $\bar{\phi}_{\rm lin}\, V/n$. The $P_c(s)$ data follow the $\sim s^{-3/2}$ scaling typical of melts and ideal-chains in three dimensions, the effective exponent from a power-law best fit to the data for $150\times150\times150$ systems being $-1.505 \pm 0.003$ (solid line). Also the mean square internal distance (inset) exhibits the expected scaling for melts, $\langle R^2(s) \rangle \propto s$, the best fit exponent being $1.00 \pm 0.01$.

For the inter-chain contact statistics, shown in Figure \ref{fig:fig4_metric}b, we computed two distinct observables for the same near monodisperse samples. First, we addressed the length dependence of the average number of inter-chain monomeric contacts, $n_{\rm contacts}$, made by a given linear chain. Next, we computed the number of overlapping chains, $n_{\rm overlap}$, i.e.~chains making at least one contact with a given one. Contributions from periodic images were included in both cases. We observed that $n_{\rm contacts}$ grows as $N^{\beta_1}$, with $\beta_1 = 0.994 \pm 0.006$ while $n_{\rm overlap}$ scales as $N^a$, with $a = 0.48 \pm 0.01$. We recall that the expected theoretical exponents for melts are $\beta_1^{\rm theor}= 1 + d (2 \nu -1)$ and $a^{\rm theor}=\nu \, d -1$, where $d$ is the dimensionality and $\nu$ is the metric exponent~\cite{halverson2014melt}. Our observed $\beta_1$ and $a$ are compatible with these theoretical expressions specialized to three-dimensional melts ($d=3$ and $\nu=1/2$), $\beta_1^{\rm theor}= 1$ and $a^{\rm theor}=1/2$.

Notice that all considered observables collapse onto the same master curves across various $n$ and $V$, even though the density of periodic chain images varies substantially across these combinations. This indicates that the metric scaling of the giant linear chains at $\bar{\phi}_{\rm lin}$ effectively lies in a universal melt regime, insensitive to microscopic lattice details, and interactions with both periodic images and the finite ring background.

Capitalizing on this, we use the giant chains equilibrated by SAMC as a gateway to access properties of linear melts at previously unexplored chain lengths and densities. Given the centrality of topological constraints for rationalizing and predicting the physical properties of melts, we next direct our analysis to intra- and inter-chain entanglement across scales.

{\bf Knotting in giant linear melts.} For the intra-chain topological entanglements, we study how the abundance and complexity of physical knots vary with the length of the macroscopic linear chains. The most straightforward setting for this analysis is to consider the $n=1$ case conditioned at $\bar{\phi}_{\rm lin}$, which yields giant chains with near monodisperse lengths, $N \simeq 0.9 V$. The physical knotted state of the (unwrapped) macroscopic linear chains was established by joining the ends with minimally-invasive auxiliary arcs and then evaluating topological invariants, see Methods section. Because this topological profiling is more costly than SAMC sampling, we performed it up to $N \simeq 1.1 \times 10^5$, corresponding to systems of size $50 \times 50\times 50$.

\begin{figure*}[t!]
\centering
\includegraphics[width=\textwidth]{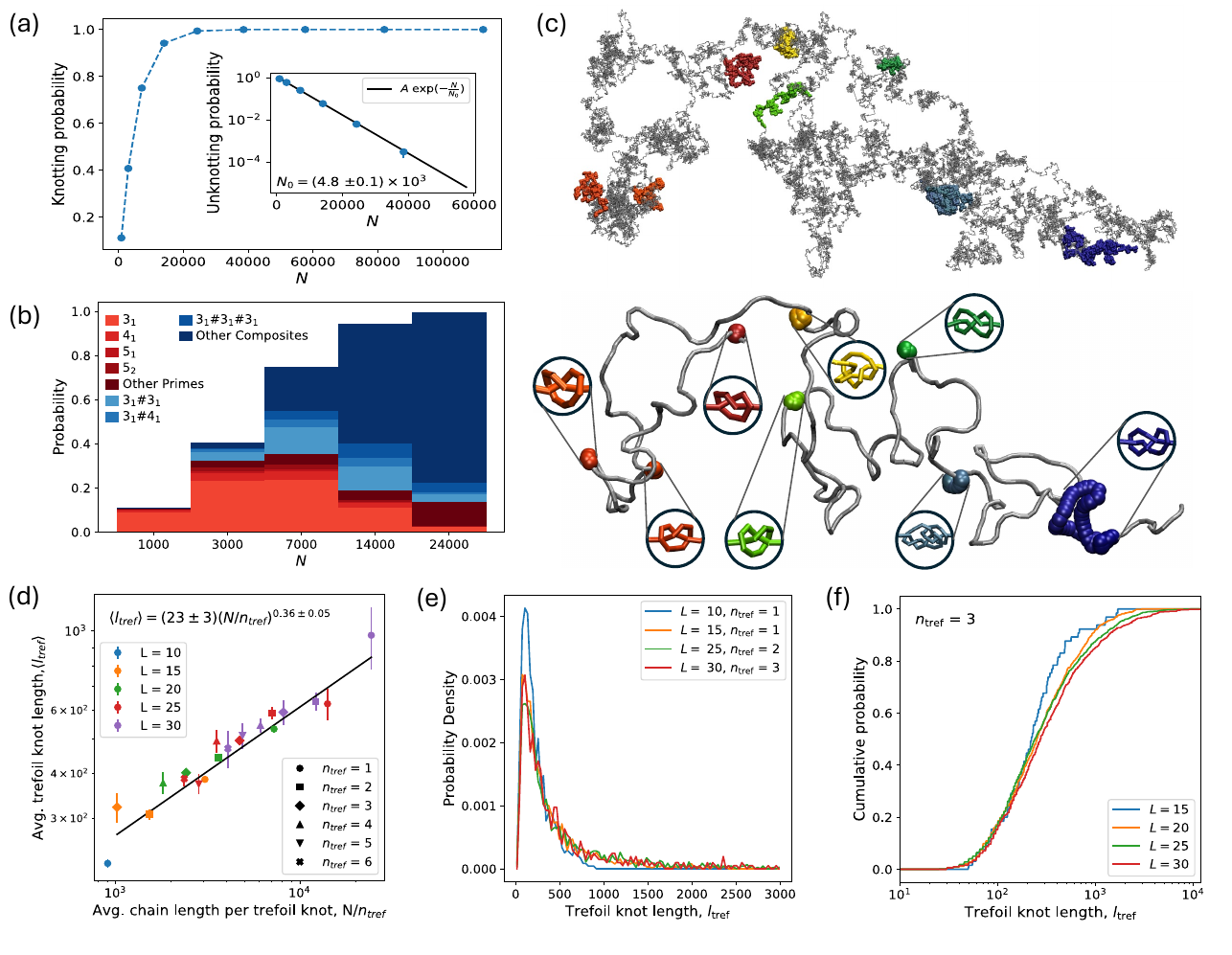}
\caption{{\bf Intra-chain entanglement: physical knotting of giant linear chains.} (a) Knotting probability of giant chains versus their length $N$. The complementary unknotting probability is shown in the inset along with the exponential best fit to the data. (b) Stacked histogram illustrating the contributions of various prime and composite knots to the overall knotting probability. (c) Typical configuration (unwrapped representation) of a giant chain of $N \simeq 5.8 \times 10^4$ monomers before (top) and after (bottom) a topology-preserving geometrical simplification. In the simplified configuration, spatially separated localized prime components are clearly visible, see also close-ups in the callouts. The corresponding knotted regions are highlighted with matching colours in the original configuration. Data are based on $10^4$ independent configurations conditioned at $\bar{\phi}_{\rm lin}$ for various system sizes and $n=1$ linear chains. (d) Average trefoil knot length, $\langle l_{\rm tref}\rangle$ as a function of $N/n_{\rm tref}$, the chain contour length rescaled by the number of trefoil components. Only $(L,n_{\rm tref})$ combinations with at least 50 measurements of trefoil knot length were considered. Data for these combinations, distinguished by different colours and symbols, collapse onto a master curve that is well described by the power law best fit $\langle l_{\rm tref}\rangle \propto (N/n_{\rm tref})^{0.36\pm0.05}$ (dashed line). The outlier data point for $(L=10,n_{\rm tref}=1)$ was excluded from the fit. (e) Probability density of $l_{\rm tref}$ for various $(L,n_{\rm tref})$ combinations. The corresponding cumulative distributions for $n_{\rm tref}=3$ and various $L$ are shown in panel (f). Error bars in panels (a) and (d) represent standard errors.}
\label{fig:fig5_knots}
\end{figure*}

Figure \ref{fig:fig5_knots} shows the (un-)knotting probability of the giant chains as a function of $N$. The data indicate an effective exponential decay, $P_{\rm unknot}(N) \sim \exp(-N/N_0)$, with characteristic knotting length $N_0 = (4.8 \pm 0.1) \times 10^3$. This value lies between those of two well-studied cases. It is an order of magnitude smaller than for isolated self-avoiding chains ($N_0 \simeq 10^5$ for both open and closed chains on- and off-lattice) but an order of magnitude larger than for ideal chains ($N_0 \simeq 300$ Kuhn lengths~\cite{deguchi1997universality,katritch2000tightness} about $450$ steps in our model, see Methods). The intermediate value of $N_0$ reflects the crossover of effective interactions in dense melts, where excluded volume is screened at large scales while it persists locally. For completeness, we note that the observed $N_0$ for the giant melts is also much larger than for self-avoiding polymers below the $\theta$-point. In this regime, corresponding to bad solvent conditions, polymer chains are individually collapsed and the reported characteristic knotting lengths of these dense and hence highly entangled globules, which have been studied for circular chains~\cite{van1990knot,virnau2005knots,baiesi2014knotted}, are of the order of $10^2-10^3$. Consequently, melt chains are substantially underknotted relative to pure random walks~\cite{meyer2018mapping} and collapsed polymers, while also being overknotted compared to isolated self-avoiding chains.

The knot spectrum, which we could resolve near exhaustively for lengths up to $N \simeq 2.4 \times 10^4$, is dominated by composite knots, which are connected sums of prime components (Figure \ref{fig:fig5_knots} and Supplementary Table 2).
A sliding-window scan along the chain contour reveals that for $N \gg N_0$ these composites typically manifest as multiple localized simple prime knots separated by long unknotted segments (Figure \ref{fig:fig5_knots}c). The observed spatial factorization is nontrivial, since the multiple prime components could instead coalesce in a single delocalized tangle.

\begin{figure*}[t!]
\centering
\includegraphics[width=\textwidth]{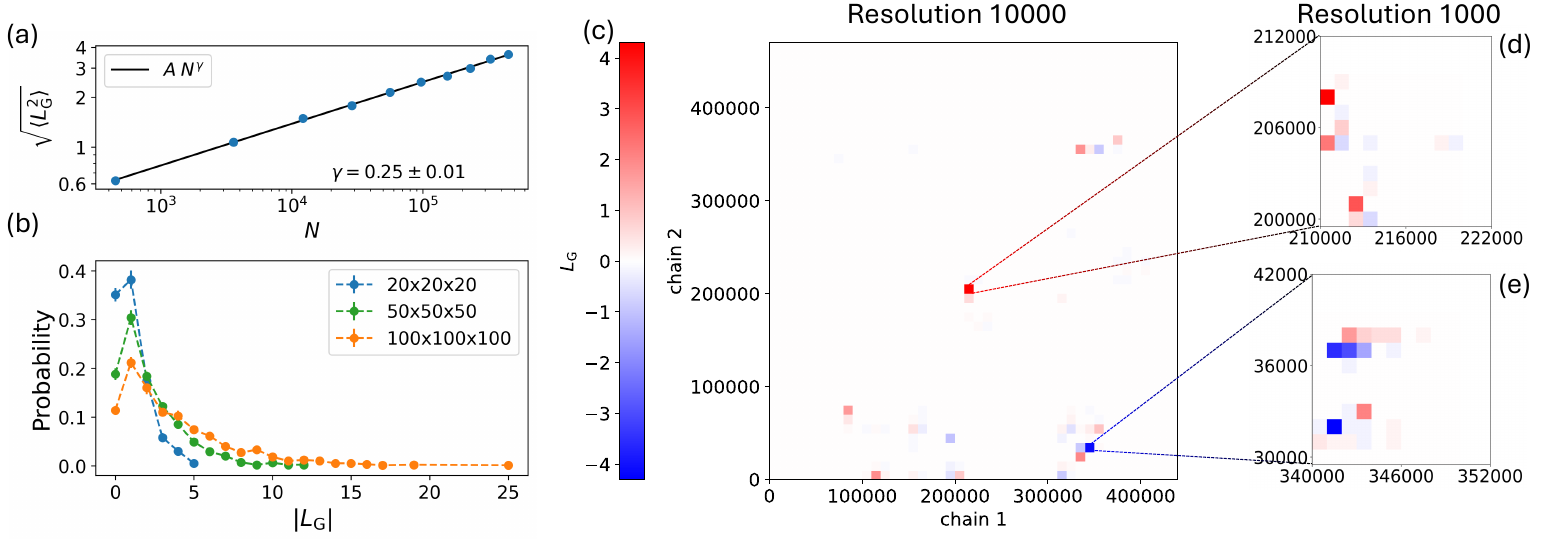}
\caption{{\bf Inter-chain entanglement: physical linking of giant linear chains.} (a) Root-mean-square Gauss linking integral $\sqrt {\langle L_{\rm G}^2 \rangle }$ versus chain length $N$. The solid line is a power-law best fit to the data; the exponent is indicated. (b) Histogram of the $|L_{\rm G}|$ probability distribution, with bins centred at integer values. Data are for $10^3$ independent configurations conditioned at $\bar{\phi}_{\rm lin}$ for various system sizes and $n=2$ near-monodisperse linear chains. (c) Typical heatmap $M_{L_{\rm G}}$ of local segment-segment contributions to the Gauss linking integral of two contacting (unwrapped) giant chains in a $100\times 100\times 100$ system. The heatmap is computed at the resolution of $10^4$ monomers per segment. Finer resolution close-ups are shown in panels (d,e). Error bars in panels (a) and (b) represent standard errors; those in (a) are smaller than the symbols.}
\label{fig:fig6_lk}
\end{figure*}

Motivated by the abundance of trefoils ($3_1$) among prime knots and prime components (Supplementary Table 2) we analyzed their average contour length, $\langle l_{\rm tref} \rangle$. To this end, we isolated individual trefoil-knotted components with a bottom-up scan of the chains (Supplementary Note 8). Figure \ref{fig:fig5_knots}d shows that $\langle l_{\rm tref} \rangle$ depends on both the overall chain length, $N$, and the number of trefoil-knotted components, $n_{\rm tref}$, accommodated by the chain. Except for the smallest system ($L=10$), the $\langle l_{\rm tref}(N,n_{\rm tref})\rangle$ data points fall on the same master curve when plotted versus the reduced chain length $N/n_{\rm tref}$. The master curve is well described by the power law (dashed line in Figure \ref{fig:fig5_knots}d)
\begin{equation}
\langle l_{\rm tref}\rangle \propto \left(\frac{N}{n_{\rm tref}}\right)^{0.36 \pm 0.05} .
\end{equation}
\noindent Several observations are in order. First, the result is qualitatively consistent with previous studies of single self-avoiding open and closed chains, which found that the average contour length of single prime knots scaled sublinearly with chain length, indicating weak knot localization~\cite{farago2002pulling,marcone2005length,virnau2005knots,mansfield2010properties,tubiana2013spontaneous}.

Second, it is interesting to compare the observed exponent with previous estimates for trefoil-knotted polymers in the absence of attractive interactions. Reported values span a broad range, possibly also due to the limited length of the studied chains, as noted by Mansfield and Douglas~\cite{mansfield2010properties}. Specifically, previous estimates include $0.74 \pm 0.14$ and $0.54$ (\cite{marcone2005length} and~\cite{mansfield2010properties}, respectively) for lattice rings, $0.63$ for a polyethylene model~\cite{virnau2005knots}, $0.4\pm0.1$ as inferred from the mechanical response of chains stretched between two plates~\cite{farago2002pulling}, and $0.44\pm0.08$ for fully-flexible off-lattice chains of up to $1.5\times 10^4$ monomers~\cite{tubiana2013spontaneous}. The value found here for the giant chains in a melt, $0.36 \pm 0.05$, is at the lower end of the above range, and is closest to the last two estimates.

Third, our results show that this scaling extends to the individual localized prime components of composite knots, with $N/n_{\rm tref}$ emerging as the appropriate scaling variable to account for the competition of the multiple components that share the same chain contour.

Finally, further insight into the characteristic knot length emerges from the probability distributions of the individual trefoil-knot lengths, $l_{\rm tref}$, shown in Figure \ref{fig:fig5_knots}e,f for various $(N,n_{\rm tref})$ combinations. In qualitative accord with Ref.~\cite{tubiana2013spontaneous}, the data show that: (i) the distributions are unimodal, (ii) they are peaked at about $\bar{l}_{\rm tref}\sim 100$ almost independently of $N$ and $n_{\rm tref}$, and (iii) have a slowly decaying tail at large knot lengths. At the same time, the tails of the distributions decay slowly enough that a non-negligible portion of the probability density extends to the largest observed knot lengths (Figure \ref{fig:fig5_knots}f). Therefore, while the modal trefoil-knot length is independent of $N$ and $n_{\rm tref}$, the average one depends on $N/n_{\rm tref}$ because increasing this rescaled length extends the cutoff of the slowly decaying probability tail.

As a complement to the above considerations on the length and localization of trefoil knots, we discuss a minimal stochastic model for their multiplicity in composite knots. To this end we consider a Poisson insertion process in which localized trefoils are placed independently along the chain contour. Inspired by theoretical arguments for isolated chains~\cite{sumners1988knots,pippenger1989knots,van1990knot,grosberg1997disordered}, we take the insertion rate as $\bar{N}_{\rm tref}^{-1}$, where $\bar{N}_{\rm tref} \simeq (5.0 \pm 0.2)\times 10^3$ is the chain length corresponding to the peak abundance of trefoils (Figure~\ref{fig:fig5_knots} and Supplementary Figure 10). This process, despite underestimating total knotting by neglecting the insertion of other prime types, yields $P_{\rm unknot}(N) \sim \exp(-N/\bar{N}_{\rm tref})$, capturing both the observed exponential decay and the correct order of magnitude for $N_0$. Thus, the model ties both results to the observed spatial localization of the simple prime components.

{\bf Linking in giant linear melts.} For the inter-chain topological entanglement, we investigate the pairwise physical linking of contacting giant chains, studying its localization and dependence on chains' length. To achieve, again, near monodisperse conditions, we considered the case $n=2$ conditioned at $\bar{\phi}_{\rm lin}$, further restricting the dataset to configurations with a $<10\%$ difference between the two chain lengths, hence equal to $N_1 \simeq N_2 \simeq 0.45\,V$.
To quantify the topological interlocking of the two chains, we computed the Gauss linking integral $L_{\rm G}$ (Supplementary Note 9) of their closest periodic images, defined as the unwrapped images with the smallest center-of-mass separation (see Methods).

 Because the ensemble average $\langle L_{\rm G} \rangle $ vanishes by symmetry, we characterized the entanglement magnitude using the root-mean-square and mean absolute values, $\sqrt{\langle L_{\rm G}^2 \rangle}$ and $\langle |L_{\rm G}| \rangle$. Note that for linear chains $L_{\rm G}$ is generally not an integer. The analysis was performed up to $N \sim 4.5\times10^5$ monomers, corresponding to systems of size $100 \times 100\times 100$.

Figure \ref{fig:fig6_lk}a illustrates the linking magnitude as a function of $N$. Intuitively, one might expect that tightly contacting chains in densely packed systems should form tangles of complexity rapidly growing with $N$. In stark contrast, the linking magnitude increases only slowly with chain length,
\begin{equation}
\sqrt{\langle L_{\rm G}^2 \rangle} \sim N^\gamma , \ \ \  \gamma \simeq 0.25 \pm 0.01 \ ,
\end{equation}
and remains modest even for our longest chains. Indeed, the most probable value of $|L_{\rm G}|$ remains of order unity throughout the entire explored length range (Figure \ref{fig:fig6_lk}b).

To resolve the spatial organization of the overall weak interlockings, we decomposed $L_{\rm G}$ into segment-segment contributions (Methods and Supplementary Note 9). The heatmaps $M_{L_{\rm G}}$ constructed through this double-contour scan are remarkably sparse even for the longest chains, as illustrated in Figure \ref{fig:fig6_lk}c-e for a representative $M_{L_{\rm G}}$. At the segment resolution of $\Delta=10000$ monomers, only about $20$ segment pairs contribute significantly to the overall linking integral ($L_{\rm G} \simeq 1.19$). Importantly, these contributions appear with both signs. Using finer segment subdivisions reveals that the significant entries correspond to localized clusters.

We conclude that, despite the overall tight packing, nearby giant chains have sparse and localized entanglements. Even at the largest lengths, pairwise chain interlockings occur over only a few and widely separated distinct regions.

Both this linking factorization and the observed exponent $\gamma\simeq 0.25 \pm 0.01$ can be rationalized and connected by a simple scaling argument for monodisperse melts at fixed density. A given chain overlaps with $n_{\rm overlap} \sim \sqrt{N}$ neighbouring ones (see also Figure \ref{fig:fig4_metric}b), with the total number of interlockings scaling as $\sim N/L_e$, where $L_e$ is the entanglement length. Thus, the typical number of distinct interlockings between two neighbouring chains scales as $n_{\rm links}^{\rm pair} \sim N/(L_e\, n_{\rm overlap}) \propto N^{1/2}$. Since these topological constraints are sparse and independent, they contribute to $L_{\rm G}$ with random (unbiased) signs. Resorting to the central limit theorem, the argument gives,
\begin{equation}
\sqrt{\langle L_{\rm G}^2 \rangle^{\rm theor}} \sim \sqrt{n_{\rm links}^{\rm pair}} \propto N^{1/4} \ ,
\end{equation}
\noindent thus accounting for both the observed scaling exponent and the sparsity of the $M_{L_{\rm G}}$ heatmaps.

As a complement, we note that the $L_{\rm G}$ distributions have fatter than Gaussian tails (Supplementary Figure 12), suggesting long-range correlations between interlockings, a feature warranting future investigation.

\section*{Discussion}
We introduced a self-assembly Monte Carlo (SAMC) scheme for equilibrating fully packed ring-linear blends on a lattice via strictly local endpoint bond swaps. The method samples an ensemble in which the number of linear chains, $n$, is conserved while chain lengths and ring statistics fluctuate.

This allows SAMC to bypass the bottlenecks of conventional fixed-composition schemes, achieving a linear scaling of the global decorrelation time with system size, $\tau_{\rm eq}\sim V^{1.001 \pm 0.002}$. This scaling inverts the usual hierarchy of computational costs, shifting it from the collection of independent states to their analysis. Accordingly, we equilibrated fully packed blends in which a few linear chains collectively comprise $\sim 9.7\times 10^{8}$ monomers, and computed complex observables for individual chains of length up to $N \sim 10^{6}$.

In the SAMC stationary ensemble, macroscopic linear components emerge spontaneously, collectively occupying $\simeq 90$\% of the volume. At the modal linear volume fraction, $\bar{\phi}_{\rm lin}\simeq 0.9$, the $n$ linear chains are statistically equivalent and individually extensive, while the complementary ring background remains non-extensive and dominated by short loops. As a result, the giant chains reproduce the metric statistics of dense linear melts, placing the large-scale behaviour accessed by SAMC in the melt universality class.

We harnessed SAMC to characterize topological entanglement in melts at packing density and chain lengths up to $N \sim 10^6$, a regime typically inaccessible to equilibrium sampling. For intra-chain entanglement, i.e.~physical knotting, the unknotting probability is well described by an exponential decay with characteristic length $N_{0}\simeq 4.8\times 10^{3}$, substantially smaller than in dilute self-avoiding conditions. In addition, the growth of topological complexity with $N$ is driven by the spatial factorization into localized prime knots separated by extended unknotted segments, similar to a Poisson insertion process.

For inter-chain entanglement, i.e.~physical linking, we observed an analogous spatial factorization of the interlocking between contacting chain pairs. The Gauss linking integral, $L_{\rm G}$, remains modest and of order unity even at the longest analyzed lengths, $N\sim 4.5\times 10^5$, and its magnitude $\sqrt{\langle L_{\rm G}^2\rangle}$ grows only as $N^{1/4}$. Analysis of segment-segment contributions to $L_{\rm G}$ shows that this weak growth arises from sparse and spatially localized interlockings. A simple scaling argument based on the number of such topological constraints and their unbiased random-sign contribution to $L_{\rm G}$ rationalizes both the exponent and the observed sparsity.

The performance of SAMC, which is also straightforward to implement, makes it well suited to tackle strongly entangled polymer systems at lengths where conventional equilibration is typically unfeasible. A standing open problem addressable with SAMC is the characterization of multi-chain entanglements~\cite{barkataki2022jones,barkataki2024virtual}, which remain poorly understood and, e.g., may account for the interlocking correlations suggested by the non-Gaussian tails of the measured $|L_{\rm G}|$ distributions. Spatial confinements relevant to experimental setups (channels, slits and cavities) can be accessed with suitable embedding geometries and open boundaries (Supplementary Note 2). The microscopic details of the model can be refined too. The resolution can be enhanced with higher-coordination lattices and bending rigidity can be treated with the same local updates supplemented by a Metropolis acceptance criterion. More generally, SAMC-generated configurations can be used, after off-lattice relaxation, as equilibrated initial states for molecular dynamics simulations of large dense polymer systems. This would enable studying the manifestations and implications of topological constraints across scales, a central problem from designed materials, such as slide-ring polymer networks, to biological soft matter, such as eukaryotic chromosomes.

In principle, SAMC could be extended to control the dispersity of the linear chains. For instance, the original SAMC could be used to sample equilibrated configurations with a substantial number of distinct chains, e.g.~$n=O(\sqrt{V})$, thus ensuring a non-giant average chain length, $O(\sqrt{V})$. To achieve near-monodispersity, these configurations could then be separately evolved with a modified SAMC in which moves are accepted or rejected with a Metropolis-like criterion favouring small chain-length dispersity. This additional evolution, requiring backbone tracing at each step, would steer the original configurations towards near-monodispersity. Among other applications, this would enable more direct comparisons with standard ring-linear blends and linear melts.

A further interesting future direction is to extend SAMC beyond the fully packed regime. To this end, one could equip SAMC with additional moves, e.g.~attaching bonds to empty lattice sites while deleting terminal bonds to conserve density. Though the sampling efficiency would expectedly degrade as the packing fraction is lowered, with such an extension it might become possible to study how knotting and linking statistics cross over from dense melts to dilute self-avoiding polymers.

SAMC could also be generalized to simulate blends of living polymers~\cite{cates1990statics} by supplementing the present dynamics with scission and recombination moves that create or anneal nearby pairs of free ends. More specifically, the number of linear chains, $n$, could be allowed to fluctuate by removing non-terminal bonds through split moves that increment $n$ by 1, and by bridging nearest-neighbour endpoints through merge moves that decrease $n$ by 1. This would allow for sampling living polymer blends in which the linear chains fluctuate in both number and length, as the circular chains in the background already do. Though the efficiency of such extensions would need to be determined, it would be interesting to use the approach to study how reversible breakage and recombination modify the equilibrium properties of dense ring-linear blends.

\section*{Methods}
{\bf Model and initialization.} Our approach is inspired by recent work that recasts the sampling of dense polymer systems as a binary optimization (QUBO) problem amenable to quantum annealing machines~\cite{micheletti2021polymer,slongo2023quantum,slongo2025computing,Grosberg2025}. The QUBO representation is based on binary variables, corresponding to occupation states of lattice sites, edges, and corner turns, that interact via local bonding rules but without global connectivity constraints. This motivated our modelling of ring-linear polymer blends in the self-assembly ensemble rather than at fixed composition.

The initial space-filling SAMC state for $n$ linear chains and the ring background on $L\times L\times L$ periodic simple-cubic lattice can be conveniently prepared with the following two-step procedure.
First, one generates a deterministic Hamiltonian path that, starting from a corner of the $L\times L \times L$ box, fills successive layers in a serpentine fashion. Removing $n-1$ non-consecutive bonds from the obtained path yields the desired initial space-filling configuration with $n$ open chains, with an initial ring count of zero.
The procedure can be used with all admissible values of $n$, from 1 to $n_{\rm max}=\lfloor (V-1)/2 \rfloor$. Excluding the $n=V/2$ case, when admissible by lattice parity constraints, is necessary because it corresponds to configurations entirely filled by chains of length 1, which are blocked under the SAMC evolution (Supplementary Note 2).
Under the subsequent Monte Carlo dynamics, the number of linear chains, $n$, is conserved while chain lengths and ring counts vary. The tests of uniformity and ergodicity of the self-assembly Monte Carlo, including a statistical analysis after Ref.~\cite{mansfield2006unbiased}, are presented in Supplementary Figure 1.

The nominal Kuhn length of polymers on simple-cubic lattices can be estimated by considering non-backtracking random walks, for which the mean scalar product of two consecutive unit-length steps is $\langle c \rangle =1/5$. Using the Kuhn length expression for freely-rotating chains, $(1+c)/(1-c)$, and replacing $c$ with $\langle c \rangle$ yields a nominal value of $3/2$.

{\bf Global decorrelation time.} To quantify global decorrelation we use the normalized edge-state overlap $q$ between two configurations $\Gamma_0$ and $\Gamma_1$:
\begin{equation}
 q(\Gamma_0,\Gamma_1) = \frac{n_{\rm eq}(\Gamma_0,\Gamma_1)}{E}\,,
\end{equation}
where $n_{\rm eq}$ counts the lattice edges with identical state (empty or occupied) in both configurations and the normalizing factor $E = 3V$ is the total number of edges.

The normalized autocorrelation function for $q$ is:
\begin{equation}
C(t) = (\langle q(t) \rangle - \bar{q} )/ (1 - \bar{q})\ ,
\end{equation}
\noindent where $\langle q(t) \rangle$ is the average overlap of configurations at time lag $t$, measured in Monte Carlo steps, and $\bar{q}$ is the average overlap of independent configurations, estimated numerically (Supplementary Figure 3).
The global decorrelation time, $\tau_{\rm eq}$, is obtained by integrating $C(t)$ from $t=0$ up to the first time lag when $C(t) < 10^{-2}$.

{\bf Statistical analysis of observables.} Power-law best fits were performed via linear regression of the log-log data. The reported errors of the exponents correspond to the $3\sigma$ confidence interval. For each combination of system size and $n$, data in Figure \ref{fig:fig4_metric}a involve a double average. First, for each considered giant chain, we compute the contact probability of all pairs of monomers at separation $s$. Next, we compute $P_c(s)$ as the mean of these individual contact probabilities at fixed $s$. In the plots of Figures~\ref{fig:fig4_metric}b, \ref{fig:fig5_knots}, \ref{fig:fig6_lk}, where data for near-monodisperse chains are presented as a function of chain length, the reported value of $N$ corresponds to $\bar{\phi}_{\rm lin}\,V/n$, i.e.~the reference length around which 10\% length fluctuations are allowed. Statistical errors on the knotting probabilities were obtained from block analysis, using $m=5$ or $m=10$ blocks per dataset depending on data numerosity.

{\bf Detection of physical knots.} The knotting probability and knot spectrum of the giant linear chains were established by computing topological invariants on their circularized versions, obtained by applying the minimally invasive closure procedure of Ref.~\cite{tubiana2018kymoknot}. For a given open chain, the latter involves comparing the termini separation, $D_{\rm TT}=d_{{\rm t1},{\rm t2}}$, with the sum of their distances to the chain convex hull, $D_{\rm TCH}=d_{{\rm t1},{\rm ch}}+d_{{\rm t2},{\rm ch}}$. If $D_{\rm TT}\le D_{\rm TCH}$, the termini are bridged directly; otherwise, each terminus is prolonged along its distance vector to the convex hull by a length ten times the chain gyration radius and the prolonged ends are then connected. Choosing the smaller of $D_{\rm TT}$ and $D_{\rm TCH}$ is a practical way to limit spurious entanglement from closure arcs by minimizing the contour length of the portion inside the convex hull, which can topologically interfere with the open chain.
To speed up the calculation of topological invariants for the giant chains, we apply topology-preserving geometrical simplifications, first on-lattice using BFACF-type moves~\cite{berg1981random,de1983polymers} and then off-lattice with KMT-type moves~\cite{koniaris1991self,taylor2000deeply}. After simplification, we compute Alexander determinants and Dowker codes. After algebraic simplification and factorization, the latter are compared against lookup tables of prime knots with up to 16 crossings, allowing near complete ($>95$\%) determination of the knot spectrum for giant chains with up to $N\simeq 2.4\times 10^4$ monomers. For longer chains we conservatively discriminate between knotted and unknotted states using the determinants of the Alexander polynomial. Applying the same knot-profiling procedure to systematic bottom-up scans of chain segments of increasing length enables localizing the knotted components. Details of the scanning procedure and knot spectrum, including a best-fit analysis of the trefoil-knot probability after Ref.~\cite{deguchi1994statistical}, are provided in Supplementary Note 8, Supplementary Table 2 and Supplementary Figure 10.

{\bf Detection of physical links.} To quantify the degree of entanglement of two giant linear chains, we compute the Gauss linking integral $L_{\rm G}$ of their closest unwrapped periodic images, i.e.~the pair of images with the minimal center-of-mass separation. Except for accidental degeneracies, the distance-minimizing pair is unique up to common translations by integer linear combinations of the simulation-box basis vectors, under which $L_{\rm G}$ is invariant. The latter was evaluated using the method detailed in Supplementary Note 9.
The direct evaluation of $L_{\rm G}$ involves $O(N^2)$ elementary operations. Applying on- and off-lattice topology-preserving simplifications afforded an order-of-magnitude speed-up while leaving $L_{\rm G}$ unchanged to within $1$\% (Supplementary Figure 11). To establish the degree of localization of physical links we subdivide each chain into equally long segments and compute the segment-segment contributions to $L_{\rm G}$, yielding matrices $M_{L_{\rm G}}$ capturing the degree of physical interlocking between the various regions of the two chains.

\section*{Data availability}
The data generated in this study and presented in the main text and Supplementary Information plots have been deposited together with the corresponding Jupyter notebooks in the Zenodo repository under accession code \href{https://doi.org/10.5281/zenodo.20504264}{10.5281/zenodo.20504264}\cite{data_repository}.

\section*{Code availability}
The source code of SAMC has been deposited in the GitHub repository under accession code \href{https://github.com/FrancescoSlongo/SAMC_Self_assembly_Monte_Carlo}{FrancescoSlongo\/SAMC\_Self\_assembly\_Monte\_Carlo}\cite{code_repository}. Compilation requires users to provide their own licensed copy of the specified random number generator.

\normalem
\bibliographystyle{achemso}
\bibliography{bibliography}

@article{deutsch1997long,
  title={Long range moves for high density polymer simulations},
  author={Deutsch, JM},
  journal={The Journal of chemical physics},
  volume={106},
  number={21},
  pages={8849--8854},
  year={1997},
  publisher={American Institute of Physics}
}

@article{rosenbluth1955monte,
  title={"Monte Carlo calculation of the average extension of molecular chains"},
  author={Rosenbluth, Marshall N and Rosenbluth, Arianna W},
  journal={The Journal of Chemical Physics},
  volume={23},
  number={2},
  pages={356--359},
  year={1955},
  publisher={American Institute of Physics}
}

@article{tesi1996monte,
  title={"Monte Carlo study of the interacting self-avoiding walk model in three dimensions"},
  author={Tesi, M.C. and Van Rensburg, E.J. Janse and Orlandini, E. and Whittington, S.G.},
  journal={Journal of statistical physics},
  volume={82},
  number={1},
  pages={155--181},
  year={1996},
  publisher={Springer}
}

@article{clisby2010accurate,
  title={Accurate estimate of the critical exponent $\nu$ for self-avoiding walks via a fast implementation of the pivot algorithm},
  author={Clisby, N.},
  journal={Physical review letters},
  volume={104},
  number={5},
  pages={055702},
  year={2010},
  publisher={APS}
}

@article{grassberger1997pruned,
  title={Pruned-enriched Rosenbluth method: Simulations of $\theta$ polymers of chain length up to 1 000 000},
  author={Grassberger, P.},
  journal={Physical Review E},
  volume={56},
  number={3},
  pages={3682},
  year={1997},
  publisher={APS}
}

@article{smrek2013novel,
  title={A novel family of space-filling curves in their relation to chromosome conformation in eukaryotes},
  author={Smrek, Jan and Grosberg, Alexander Y},
  journal={Physica A: Statistical Mechanics and its Applications},
  volume={392},
  number={24},
  pages={6375--6388},
  year={2013},
  publisher={Elsevier}
}

@article{svaneborg2023multiscale,
  title={Multiscale equilibration of highly entangled isotropic model polymer melts},
  author={Svaneborg, Carsten and Everaers, Ralf},
  journal={The Journal of Chemical Physics},
  volume={158},
  number={5},
  year={2023},
  publisher={AIP Publishing}
}

@article{micheletti2021polymer,
  title = {Polymer Physics by Quantum Computing},
  author = {Micheletti, Cristian and Hauke, Philipp and Faccioli, Pietro},
  journal = {Phys. Rev. Lett.},
  volume = {127},
  issue = {8},
  pages = {080501},
  numpages = {7},
  year = {2021},
  month = {Aug},
  publisher = {American Physical Society},
  doi = {10.1103/PhysRevLett.127.080501},
  url = {https://link.aps.org/doi/10.1103/PhysRevLett.127.080501}
}

@article{auhl2003equilibration,
  title={Equilibration of long chain polymer melts in computer simulations},
  author={Auhl, Rolf and Everaers, Ralf and Grest, Gary S and Kremer, Kurt and Plimpton, Steven J},
  journal={The Journal of chemical physics},
  volume={119},
  number={24},
  pages={12718--12728},
  year={2003},
  publisher={American Institute of Physics}
}

@article{everaers2017flory,
  title={Flory theory of randomly branched polymers},
  author={Everaers, R. and Grosberg, A.Y. and Rubinstein, M. and Rosa, A.},
  journal={Soft matter},
  volume={13},
  number={6},
  pages={1223--1234},
  year={2017},
  publisher={Royal Society of Chemistry}
}

@article{lua2004fractal,
  title={Fractal and statistical properties of large compact polymers: a computational study},
  author={Lua, Rhonald and Borovinskiy, Alexander L and Grosberg, Alexander Yu},
  journal={Polymer},
  volume={45},
  number={2},
  pages={717--731},
  year={2004},
  publisher={Elsevier}
}

@article{mansfield2006unbiased,
  title={Unbiased sampling of lattice Hamilton path ensembles},
  author={Mansfield, Marc L},
  journal={The Journal of chemical physics},
  volume={125},
  number={15},
  pages={154103},
  year={2006},
  publisher={American Institute of Physics}
}

@article{orlandini2021topological,
  title={Topological and physical links in soft matter systems},
  author={Orlandini, Enzo and Micheletti, Cristian},
  journal={Journal of Physics: Condensed Matter},
  volume={34},
  number={1},
  pages={013002},
  year={2021},
  publisher={IOP Publishing}
}

@article{halverson2014melt,
  title={From a melt of rings to chromosome territories: the role of topological constraints in genome folding},
  author={Halverson, Jonathan D and Smrek, Jan and Kremer, Kurt and Grosberg, Alexander Y},
  journal={Reports on Progress in Physics},
  volume={77},
  number={2},
  pages={022601},
  year={2014},
  publisher={IOP Publishing}
}

@article{ma8018624,
author = {Harmandaris, Vagelis A. and Kremer, Kurt},
title = {Dynamics of Polystyrene Melts through Hierarchical Multiscale Simulations},
journal = {Macromolecules},
volume = {42},
number = {3},
pages = {791-802},
year = {2009}
}

@article{micheletti2011polymers,
  title={Polymers with spatial or topological constraints: Theoretical and computational results},
  author={Micheletti, Cristian and Marenduzzo, Davide and Orlandini, Enzo},
  journal={Physics Reports},
  volume={504},
  number={1},
  pages={1--73},
  year={2011},
  publisher={Elsevier}
}

@article{slongo2023quantum,
  title={Quantum-inspired encoding enhances stochastic sampling of soft matter systems},
  author={Slongo, Francesco and Hauke, Philipp and Faccioli, Pietro and Micheletti, Cristian},
  journal={Science Advances},
  volume={9},
  number={43},
  pages={eadi0204},
  year={2023},
  publisher={American Association for the Advancement of Science}
}

@article{slongo2025computing,
  title={Computing canonical averages with quantum and classical optimizers: Thermodynamic reweighting for QUBO models of physical systems},
  author={Slongo, Francesco and Micheletti, Cristian},
  journal={Physical Review Research},
  volume={7},
  number={2},
  pages={023116},
  year={2025},
  publisher={APS}
}

@article{prellberg2004flat,
  title={Flat Histogram Version of the Pruned and Enriched Rosenbluth Method},
  author = {Prellberg, Thomas and Krawczyk, Jaros\l{}aw},
  journal = {Phys. Rev. Lett.},
  volume = {92},
  issue = {12},
  pages = {120602},
  numpages = {4},
  year = {2004},
  publisher = {American Physical Society},
  doi = {10.1103/PhysRevLett.92.120602},
  url = {https://link.aps.org/doi/10.1103/PhysRevLett.92.120602}
}

@article{lua2005under,
  title={Under-knotted and over-knotted polymers: 2. Compact self-avoiding loops},
  author={Lua, Rhonald C and Moore, Nathan T and Grosberg, Alexander Y},
  journal={Physical and Numerical Models in Knot Theory: Including Applications to the Life Sciences},
  pages={385--398},
  year={2005},
  publisher={World Scientific}
}

@article{quitmann2023macroscopic,
  title={Macroscopic loops in the 3 d double-dimer model},
  author={Quitmann, Alexandra and Taggi, Lorenzo},
  journal={Electronic Communications in Probability},
  volume={28},
  pages={1--12},
  year={2023},
  publisher={The Institute of Mathematical Statistics and the Bernoulli Society}
}

@article{siepmann1992configurational,
  title={"Configurational bias Monte Carlo: a new sampling scheme for flexible chains"},
  author={Siepmann, J{\"o}rn Ilja and Frenkel, Daan},
  journal={Molecular Physics},
  volume={75},
  number={1},
  pages={59--70},
  year={1992},
  publisher={Taylor \& Francis}
}

@misc{grosberg2025,
  author       = {Alexander Y. Grosberg},
  title        = {Is quantum computing promising for soft matter problems?},
  howpublished = {Journal Club for Condensed Matter Physics, DOI: 10.36471/JCCM\_June\_2025\_02},
  month        = {June},
  year         = {2025},
  doi          = {10.36471/JCCM_June_2025_02}
}

@article{rubinstein1987discretized,
  title={Discretized model of entangled-polymer dynamics},
  author={Rubinstein, Michael},
  journal={Physical review letters},
  volume={59},
  number={17},
  pages={1946},
  year={1987},
  publisher={APS}
}

@article{hugouvieux2009amphiphilic,
  title={Amphiphilic multiblock copolymers: From intramolecular pearl necklace to layered structures},
  author={Hugouvieux, Virginie and Axelos, Monique AV and Kolb, Max},
  journal={Macromolecules},
  volume={42},
  number={1},
  pages={392--400},
  year={2009},
  publisher={ACS Publications}
}

@article{schram2018simulation,
  title={Simulation of ring polymer melts with GPU acceleration},
  author={Schram, Raoul D and Barkema, Gerard T},
  journal={Journal of Computational Physics},
  volume={363},
  pages={128--139},
  year={2018},
  publisher={Elsevier}
}

@article{ghosh2018epigenome,
  title={How epigenome drives chromatin folding and dynamics, insights from efficient coarse-grained models of chromosomes},
  author={Ghosh, Surya K and Jost, Daniel},
  journal={PLoS computational biology},
  volume={14},
  number={5},
  pages={e1006159},
  year={2018},
  publisher={Public Library of Science San Francisco, CA USA}
}

@article{mavrantzas1999end,
  title={"End-bridging Monte Carlo: A fast algorithm for atomistic simulation of condensed phases of long polymer chains"},
  author={Mavrantzas, Vlasis G and Boone, Travis D and Zervopoulou, Evangelia and Theodorou, Doros N},
  journal={Macromolecules},
  volume={32},
  number={15},
  pages={5072--5096},
  year={1999},
  publisher={ACS Publications}
}

@article{karayiannis2002novel,
  title={"A novel Monte Carlo scheme for the rapid equilibration of atomistic model polymer systems of precisely defined molecular architecture"},
  author={Karayiannis, Nikos Ch and Mavrantzas, Vlasis G and Theodorou, Doros N},
  journal={Physical review letters},
  volume={88},
  number={10},
  pages={105503},
  year={2002},
  publisher={APS}
}

@article{svaneborg2016multiscale,
  title={Multiscale approach to equilibrating model polymer melts},
  author={Svaneborg, Carsten and Karimi-Varzaneh, Hossein Ali and Hojdis, Nils and Fleck, Frank and Everaers, Ralf},
  journal={Physical Review E},
  volume={94},
  number={3},
  pages={032502},
  year={2016},
  publisher={APS}
}

@article{grosberg1997disordered,
  title={Disordered polymers},
  author={Grosberg, Aleksandr Yu},
  journal={Physics-Uspekhi},
  volume={40},
  number={2},
  pages={125},
  year={1997},
  publisher={IOP Publishing}
}

@article{sumners1988knots,
  title={Knots in self-avoiding walks},
  author={Sumners, DeWitt and Whittington, SG},
  journal={Journal of Physics A: Mathematical and General},
  volume={21},
  number={7},
  pages={1689},
  year={1988},
  publisher={IOP Publishing}
}

@article{pippenger1989knots,
  title={Knots in random walks},
  author={Pippenger, Nicholas},
  journal={Discrete Applied Mathematics},
  volume={25},
  number={3},
  pages={273--278},
  year={1989},
  publisher={Elsevier}
}

@incollection{dunweg1998molecular,
  title={Molecular dynamics simulations of polymer systems},
  author={D{\"u}nweg, Burkhard and Grest, Gary S and Kremer, Kurt},
  booktitle={Numerical methods for polymeric systems},
  pages={159--195},
  year={1998},
  publisher={Springer}
}

@article{dietz2022facile,
  title={Facile equilibration of well-entangled semiflexible bead--spring polymer melts},
  author={Dietz, Joseph D and Hoy, Robert S},
  journal={The Journal of Chemical Physics},
  volume={156},
  number={1},
  year={2022},
  publisher={AIP Publishing}
}

@book{doi1988theory,
  title={The theory of polymer dynamics},
  author={Doi, Masao and Edwards, Samuel Frederick},
  volume={73},
  year={1988},
  publisher={Oxford university press}
}

@article{mcleish2002tube,
  title={Tube theory of entangled polymer dynamics},
  author={McLeish, Tom CB},
  journal={Advances in physics},
  volume={51},
  number={6},
  pages={1379--1527},
  year={2002},
  publisher={Taylor \& Francis}
}

@article{everaers2004rheology,
  title={Rheology and microscopic topology of entangled polymeric liquids},
  author={Everaers, Ralf and Sukumaran, Sathish K and Grest, Gary S and Svaneborg, Carsten and Sivasubramanian, Arvind and Kremer, Kurt},
  journal={Science},
  volume={303},
  number={5659},
  pages={823--826},
  year={2004},
  publisher={American Association for the Advancement of Science}
}

@book{rubinstein2003polymer,
  title={Polymer physics},
  author={Rubinstein, Michael and Colby, Ralph H},
  year={2003},
  publisher={Oxford university press}
}

@article{hoy2020unified,
  title={Unified analytic expressions for the entanglement length, tube diameter, and plateau modulus of polymer melts},
  author={Hoy, Robert S and Kr{\"o}ger, Martin},
  journal={Physical review letters},
  volume={124},
  number={14},
  pages={147801},
  year={2020},
  publisher={APS}
}

@article{abadi2018entangled,
  title={Entangled polymer dynamics beyond reptation},
  author={Abadi, Maram and Serag, Maged F and Habuchi, Satoshi},
  journal={Nature Communications},
  volume={9},
  number={1},
  pages={5098},
  year={2018},
  publisher={Nature Publishing Group UK London}
}

@article{doi:10.1021/acs.macromol.8b01836,
author = {Gartner, Thomas
E. III and Jayaraman, Arthi},
title = {Modeling and Simulations of Polymers: A Roadmap},
journal = {Macromolecules},
volume = {52},
number = {3},
pages = {755-786},
year = {2019},
doi = {10.1021/acs.macromol.8b01836}
}

@article{doi:10.1021/acspolymersau.2c00049,
author = {Schmid, Friederike},
title = {Understanding and Modeling Polymers: The Challenge of Multiple Scales},
journal = {ACS Polymers Au},
volume = {3},
number = {1},
pages = {28-58},
year = {2023},
doi = {10.1021/acspolymersau.2c00049},
}

@article{bonato2022topological,
  title={Topological gelation of reconnecting polymers},
  author={Bonato, Andrea and Marenduzzo, Davide and Michieletto, Davide and Orlandini, Enzo},
  journal={Proceedings of the National Academy of Sciences},
  volume={119},
  number={44},
  pages={e2207728119},
  year={2022},
  publisher={National Academy of Sciences}
}

@article{wang2023entanglements,
  title={How entanglements determine the morphology of semicrystalline polymers},
  author={Wang, Zefan and Schaller, Mareen and Petzold, Albrecht and Saalw{\"a}chter, Kay and Thurn-Albrecht, Thomas},
  journal={Proceedings of the National Academy of Sciences},
  volume={120},
  number={27},
  pages={e2217363120},
  year={2023},
  publisher={National Academy of Sciences}
}

@article{grosberg1988role,
  title={The role of topological constraints in the kinetics of collapse of macromolecules},
  author={Grosberg, A Yu and Nechaev, Sergei K and Shakhnovich, Eugene I},
  journal={Journal de physique},
  volume={49},
  number={12},
  pages={2095--2100},
  year={1988},
  publisher={Soci{\'e}t{\'e} fran{\c{c}}aise de physique}
}

@article{grosse2025scale,
  title={Scale-free models of chromosome structure, dynamics, and mechanics},
  author={Grosse-Holz, Simon and Mirny, Leonid and Coulon, Antoine},
  journal={PNAS Nexus},
  pages={pgaf337},
  year={2025},
  publisher={Oxford University Press}
}

@article{barkataki2024virtual,
  title={The virtual spectrum of linkoids and open curves in 3-space},
  author={Barkataki, Kasturi and Kauffman, Louis H and Panagiotou, Eleni},
  journal={Journal of Knot Theory and Its Ramifications},
  volume={33},
  number={03},
  pages={2450006},
  year={2024},
  publisher={World Scientific}
}

@article{barkataki2022jones,
  title={The Jones polynomial of collections of open curves in 3-space},
  author={Barkataki, Kasturi and Panagiotou, Eleni},
  journal={Proceedings of the Royal Society A},
  volume={478},
  number={2267},
  pages={20220302},
  year={2022},
  publisher={The Royal Society}
}

@article{koniaris1991self,
  title={Self-entanglement in ring polymers},
  author={Koniaris, Kleanthes and Muthukumar, Murugappan},
  journal={The Journal of chemical physics},
  volume={95},
  number={4},
  pages={2873--2881},
  year={1991},
  publisher={American Institute of Physics}
}

@article{taylor2000deeply,
  title={A deeply knotted protein structure and how it might fold},
  author={Taylor, William R},
  journal={Nature},
  volume={406},
  number={6798},
  pages={916--919},
  year={2000},
  publisher={Nature Publishing Group UK London}
}

@article{berg1981random,
  title={Random paths and random surfaces on a digital computer},
  author={Berg, B and Foerster, D},
  journal={Physics Letters B},
  volume={106},
  number={4},
  pages={323--326},
  year={1981},
  publisher={Elsevier}
}

@article{de1983polymers,
  title={Polymers and g| $\varphi$| 4 theory in four dimensions},
  author={De Carvalho, C Aragao and Caracciolo, Sergio and Fr{\"o}hlich, Jurg},
  journal={Nuclear Physics B},
  volume={215},
  number={2},
  pages={209--248},
  year={1983},
  publisher={Elsevier}
}

@article{tubiana2024topology,
  title={Topology in soft and biological matter},
  author={Tubiana, Luca and Alexander, Gareth P and Barbensi, Agnese and Buck, Dorothy and Cartwright, Julyan HE and Chwastyk, Mateusz and Cieplak, Marek and Coluzza, Ivan and {\v{C}}opar, Simon and Craik, David J and others},
  journal={Physics reports},
  volume={1075},
  pages={1--137},
  year={2024},
  publisher={Elsevier}
}

@article{reith2010implementation,
  title={"Implementation and performance analysis of bridging Monte Carlo moves for off-lattice single chain polymers in globular states"},
  author={Reith, Daniel and Virnau, Peter},
  journal={Computer Physics Communications},
  volume={181},
  number={4},
  pages={800--805},
  year={2010},
  publisher={Elsevier}
}

@article{tubiana2018kymoknot,
  title={KymoKnot: A web server and software package to identify and locate knots in trajectories of linear or circular polymers},
  author={Tubiana, Luca and Polles, Guido and Orlandini, Enzo and Micheletti, Cristian},
  journal={The European physical journal E},
  volume={41},
  number={6},
  pages={72},
  year={2018},
  publisher={Springer}
}

@article{deguchi1997universality,
  title={Universality of random knotting},
  author={Deguchi, Tetsuo and Tsurusaki, Kyoichi},
  journal={Physical Review E},
  volume={55},
  number={5},
  pages={6245},
  year={1997},
  publisher={APS}
}

@article{katritch2000tightness,
  title={Tightness of random knotting},
  author={Katritch, Vsevolod and Olson, Wilma K and Vologodskii, Alexander and Dubochet, Jacques and Stasiak, Andrzej},
  journal={Physical Review E},
  volume={61},
  number={5},
  pages={5545},
  year={2000},
  publisher={APS}
}

@article{meyer2018mapping,
  title={Mapping onto ideal chains overestimates self-entanglements in polymer melts},
  author={Meyer, Hendrik and Horwath, Eric and Virnau, Peter},
  journal={ACS Macro Letters},
  volume={7},
  number={6},
  pages={757--761},
  year={2018},
  publisher={ACS Publications}
}

@article{deguchi1994statistical,
  title={"A statistical study of random knotting using the Vassiliev invariants"},
  author={Deguchi, Tetsuo and Tsurusaki, Kyoichi},
  journal={Journal of Knot Theory and Its Ramifications},
  volume={3},
  number={03},
  pages={321--353},
  year={1994},
  publisher={World Scientific}
}

@article{van1990knot,
  title={The knot probability in lattice polygons},
  author={Van Rensburg, EJ Janse and Whittington, Stuart G},
  journal={Journal of Physics A: Mathematical and General},
  volume={23},
  number={15},
  pages={3573},
  year={1990},
  publisher={IOP Publishing}
}

@article{virnau2005knots,
  title={Knots in globule and coil phases of a model polyethylene},
  author={Virnau, Peter and Kantor, Yacov and Kardar, Mehran},
  journal={Journal of the American Chemical Society},
  volume={127},
  number={43},
  pages={15102--15106},
  year={2005},
  publisher={ACS Publications}
}

@article{baiesi2014knotted,
  title={Knotted globular ring polymers: how topology affects statistics and thermodynamics},
  author={Baiesi, Marco and Orlandini, Enzo and Stella, Attilio L},
  journal={Macromolecules},
  volume={47},
  number={23},
  pages={8466--8476},
  year={2014},
  publisher={ACS Publications}
}

@misc{data_repository,
  title={Research data available through Zenodo at~\url{https://doi.org/10.5281/zenodo.20504264}}
}

@misc{code_repository,
  title={Source code available through GitHub at~\url{https://github.com/FrancescoSlongo/SAMC_Self_assembly_Monte_Carlo}}
}

@article{cates1990statics,
  title={Statics and dynamics of worm-like surfactant micelles},
  author={Cates, ME and Candau, SJ},
  journal={Journal of Physics: Condensed Matter},
  volume={2},
  number={33},
  pages={6869--6892},
  year={1990}
}

@article{tubiana2013spontaneous,
  title={Spontaneous knotting and unknotting of flexible linear polymers: Equilibrium and kinetic aspects},
  author={Tubiana, Luca and Rosa, Angelo and Fragiacomo, Filippo and Micheletti, Cristian},
  journal={Macromolecules},
  volume={46},
  number={9},
  pages={3669--3678},
  year={2013},
  publisher={ACS Publications}
}

@article{farago2002pulling,
  title={Pulling knotted polymers},
  author={Farago, Oded and Kantor, Yacov and Kardar, Mehran},
  journal={EPL (Europhysics Letters)},
  volume={60},
  number={1},
  pages={53--59},
  year={2002}
}

@article{mansfield2010properties,
  title={Properties of knotted ring polymers. I. Equilibrium dimensions},
  author={Mansfield, Marc L and Douglas, Jack F},
  journal={The Journal of chemical physics},
  volume={133},
  number={4},
  year={2010},
  publisher={AIP Publishing}
}

@article{marcone2005length,
  title={What is the length of a knot in a polymer?},
  author={Marcone, Boris and Orlandini, Enzo and Stella, AL and Zonta, Francesco},
  journal={Journal of Physics A: Mathematical and General},
  volume={38},
  number={1},
  pages={L15--L21},
  year={2005}
}

\section*{Acknowledgements}
We thank Enzo Orlandini, Angelo Rosa and De Witt Sumners for valuable feedback on the manuscript and Giovanni Bussi for advice on code optimization.

\section*{Funding}
This study was funded in part by the European Union - NextGenerationEU, in the framework of the PRIN Project "The Physics of Chromosome Folding" (code: 2022R8YXMR, CUP: G53D23000820006) and by PNRR Mission 4, Component 2, Investment 1.4\_CN\_00000013\_CN-HPC: National Centre for HPC, Big Data and Quantum Computing - spoke 7 (CUP: G93C22000600001). The views and opinions expressed are solely those of the authors and do not necessarily reflect those of the European Union, nor can the European Union be held responsible for them.


\clearpage
\onecolumngrid
\setcounter{figure}{0}
\setcounter{equation}{0}
\renewcommand{\thefigure}{S\arabic{figure}}
\renewcommand{\thetable}{S\arabic{table}}
\renewcommand{\theequation}{S\arabic{equation}}

\section*{SUPPLEMENTARY MATERIAL}
\twocolumngrid

\subsection{Self-assembly Monte Carlo: detailed balance and uniform stationary measure}
\label{sec:DB1}

Below we discuss detailed balance for self-assembly Monte Carlo (SAMC) on a fully packed $d$-dimensional periodic hypercubic lattice with $n$ linear chains (hence $2n$ endpoints). As detailed in the main text, the elementary move is articulated over the following steps.
\begin{itemize}
\item[(1)] {Endpoint selection.} Choose uniformly one of the $2n$ endpoints, denoted as $A$.
\item[(2)] {Neighbour selection.} Choose uniformly a non-bonded nearest neighbour of $A$, denoted $B$. Next, choose uniformly one of the bonded neighbours of $B$, denoted $C$ (one choice if $B$ is an endpoint, two otherwise). If $C$ is an endpoint, reject the move, i.e.~retain the current configuration.
\item[(3)] {Bond swap.} Replace bond $BC$ with $AB$.
\end{itemize}

Let us consider two admissible configurations or states, $\Gamma$ and $\Gamma^\prime$, connected by a local bond swap of the above type. The proposal probability of the forward move, $\Gamma \to \Gamma^\prime$, corresponding to the bond swap $BC \to AB$, can be written as a product of conditional probabilities:
\begin{equation}
q(\Gamma\to\Gamma^{\prime}) = P(A|\Gamma)\,P(B|A,\Gamma)\,P(C|A,B,\Gamma).
\label{eq:q1}
\end{equation}
\noindent $A$ is chosen uniformly among the $2n$ endpoints, so that $P(A|\Gamma)=(2 n)^{-1}$. $B$ is chosen uniformly among the $2d-1$ non-bonded nearest neighbours of $A$, so that $P(B|A,\Gamma)=(2d-1)^{-1}$.
Let us now denote by $m_\Gamma(B)$ the number of bonded neighbours of $B$ in $\Gamma$. This quantity can take only two values:
\begin{equation}
m_\Gamma(B)=
\begin{cases}
1, & \text{if } B \text{ is an endpoint},\\[4pt]
2, & \text{otherwise}.
\end{cases}
\end{equation}
Accordingly, we have
\begin{equation}
P(C|A,B,\Gamma)=\frac{1}{m_\Gamma(B)} .
\end{equation}
Substituting into Eq.~\ref{eq:q1} yields the proposal probability of the forward move:
\begin{equation}
q(\Gamma\to\Gamma^{\prime})=\frac{1}{2n}\,\frac{1}{2d-1}\,\frac{1}{m_\Gamma(B)}.
\label{eq:probforward}
\end{equation}

Retaining the previous site labelling, the proposal probability of the reverse move is:
\begin{eqnarray}
q(\Gamma^{\prime}\to\Gamma)&=&P(C|\Gamma^\prime)\,P(B|C,\Gamma^\prime)\,P(A|C,B,\Gamma^\prime) \nonumber\\
&=& \frac{1}{2n}\,\frac{1}{2d-1}\,\frac{1}{m_{\Gamma^\prime}(B)} .
\end{eqnarray}

Because the bond swap does not change the degree of $B$, this quantity is the same in the two configurations:
\begin{equation}
m_{\Gamma}(B)=m_{\Gamma^{\prime}}(B).
\end{equation}
Therefore, the forward and reverse proposal probabilities between two admissible configurations are identical:
\begin{equation}
q(\Gamma\to\Gamma^{\prime})=q(\Gamma^{\prime}\to\Gamma).
\end{equation}
Because all proposed admissible moves are accepted, the transition probabilities between distinct admissible configurations coincide with the corresponding proposal probabilities, while rejected proposals contribute to the self-loop probability $P(\Gamma,\Gamma)$. Therefore the SAMC Markov chain is reversible and satisfies detailed balance with respect to the uniform measure over the admissible configuration space. More precisely, the uniform measure applies to each connected component of configuration space under the SAMC dynamics.

\subsection{Blocked configurations}
Blocked configurations under the SAMC evolution can arise only when, independently of which endpoint is picked as site $A$, all possible choices of $C$ are endpoints. For such configurations, any proposed move is inevitably rejected. Such states would stall the SAMC evolution indefinitely and, conversely, would not be reachable from unblocked states.

Our considerations will be restricted to fully-packed periodic hypercubic lattices, where sites $A$ and $C$ are always at Manhattan distance 2. For a state to be blocked, all possible $A$ sites, which coincide with the set of all endpoints, must therefore have only endpoints as neighbours at Manhattan distance 2 (the $C$ sites). Equivalently, in blocked configurations all chain endpoints are mutual Manhattan next-nearest neighbours, and all Manhattan next-nearest neighbours of a given endpoint are endpoints themselves.

These constraints impose severe restrictions on the properties of blocked states that, as we discuss below, arise only in limiting cases not relevant for the giant-chain regime.

{\bf Bipartitionable lattices.} We first discuss the case of fully-packed periodic hypercubic lattices admitting a checkerboard bipartition. This corresponds to hypercubic lattices where each side of the parallelepiped periodic box consists of an even number of sites. Consequently, the volume $V$ is also even.
Based on the earlier consideration about Manhattan next-nearest neighbouring endpoints, a blocked state can arise either (i) when endpoints entirely occupy one sublattice only, or (ii) when endpoints entirely occupy both sublattices, hence all sites.

{\em Case (i), endpoints entirely occupy one sublattice only.} This case cannot be realized. The putative blocked configuration consists of $V/2$ sites of degree 1, namely the endpoints on one of the sublattices, and $V/2$ sites of degree 2, namely the non-endpoint sites on the opposite sublattice. Due to the bipartition, the sites of degree 2 have nearest neighbours that all have degree 1, so that their two bonded neighbours must necessarily be endpoints. Thus, all sites of degree 2 would be the midpoints of chains of length 2. The configuration would therefore consist entirely of $n=V/4$ linear chains of length 2, so that the total number of bonds is
\begin{equation}
n_b=2n=\frac{V}{2}.
\label{eq:nb1}
\end{equation}

At the same time, from the full-packing condition it must be that
\begin{equation}
n_b=V-n =\frac{3V}{4},
\label{eq:nb2}
\end{equation}
Because this contradicts Eq.~\ref{eq:nb1}, we conclude that blocked configurations with endpoints occupying entirely only one sublattice are not realizable on bipartitionable periodic hypercubic lattices.

{\em Case (ii), all sites are endpoints.} This case is realizable and corresponds to configurations made entirely of chains of length 1. Equivalently, blocked configurations on bipartitionable periodic hypercubic lattices arise when the number of linear chains is $n=V/2$. Such a setting is not relevant for the giant-chain regime.

{\bf Non-bipartitionable lattices.} We now consider fully-packed periodic hypercubic lattices that do not admit a checkerboard bipartition, a case that arises when at least one side of the periodic parallelepiped box consists of an odd number of sites. In such settings, the set of mutual Manhattan next-nearest neighbours covers the entire box. Accordingly, blocked states can arise only when all $V$ sites are endpoints.

We distinguish two cases, corresponding to the box volume $V$ being even or odd. When $V$ is odd, no blocked configuration can be realized. This is because the number of endpoints, $2n$, is even, and hence cannot match $V$.
Instead, the case of even $V$ is realizable and corresponds to the same situation as case (ii) for bipartitionable periodic hypercubic lattices, namely configurations where $n=V/2$ chains of length 1 cover the entire periodic box.

{\bf Lattices with open boundaries.} We now briefly comment on fully-packed hypercubic lattices with open boundary conditions, a setting relevant to study e.g.~spatial confinement. Suitable generalizations of the previous parity-based considerations show that, in this case too, blocked configurations can arise only when $V$ is even and all sites are endpoints ($n=V/2$), a limiting case not relevant to the emergence of giant chains.\\
It should also be noted that in systems with open boundaries the move rules need to be suitably generalized to account for the fact that available choices for site $B$ will vary depending on whether endpoint $A$ is in the bulk or near a boundary wall, edge or corner. These differences must be factored in to preserve detailed balance with respect to the uniform measure.

\begin{figure*}[t!]
\centering
\includegraphics[width=1\textwidth]{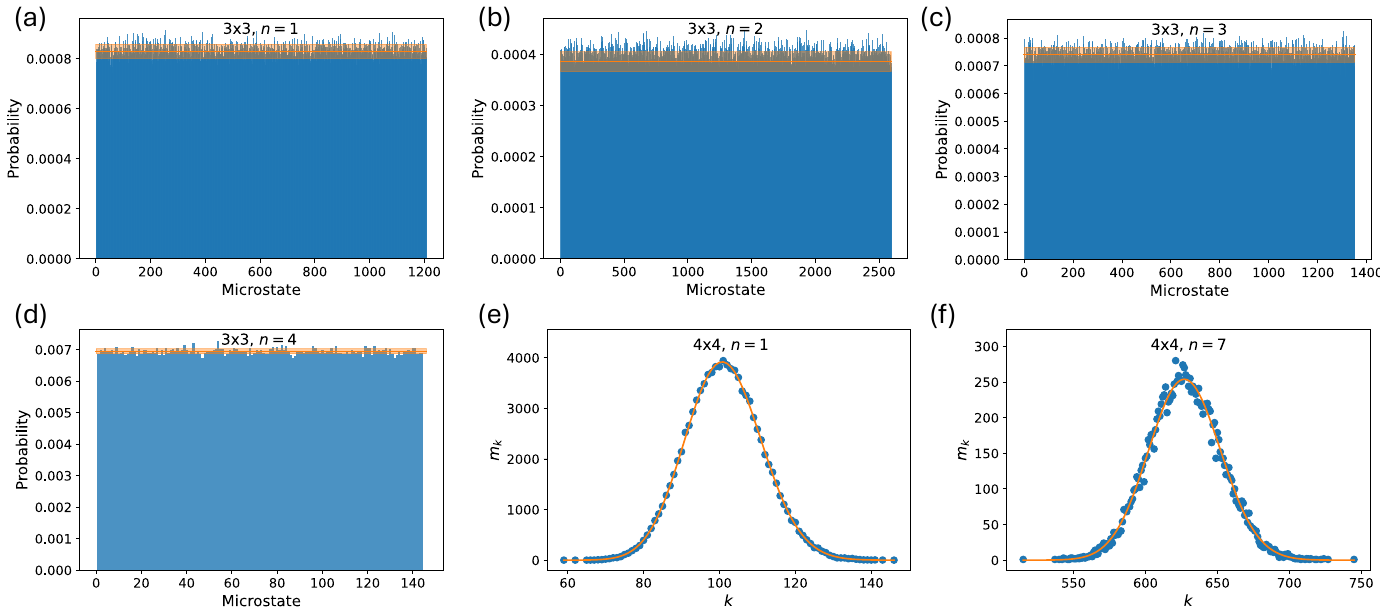}
\caption{{\bf Tests of ergodicity and uniformity of SAMC sampling on exhaustively enumerable ensembles.} Panels (a--d) show the populations of all states obtained from $10^6$ statistically independent SAMC samples for $3\times 3$ systems with $n=\{1,2,3,4\}$ linear chains, corresponding to a total of $M=\{1206,2592,1350,144\}$ states, as established by exhaustive enumeration. The horizontal orange line marks the expected mean population for uniform sampling, and the orange band indicates the corresponding standard deviation. Panel (e-f) refer to the $4\times 4$ system with $n=1$ and $n=7$ linear chains, for a total of $M=98784$  and $M=15936$ states, respectively. Due to the much larger number of possible configurations, state-by-state histograms as in panels (a--d) would be impractical to obtain and inspect. Accordingly, we instead report the quantity $m_k$, namely the number of distinct configurations sampled exactly $k$ times in $K=10^8$ statistically independent SAMC samples. The orange solid line is the corresponding binomial prediction for unbiased uniform sampling, eq.~\ref{eq:mansfield}. In all cases, the observed statistics are consistent with uniform sampling of the full enumerated configuration space.}
\label{fig:fig_SI_ergodicity}
\end{figure*}

\subsection{Effective ergodicity and uniformity.}
Detailed balance implies a uniform stationary measure on each connected component of configuration space, but does not by itself establish that the SAMC dynamics is irreducible. Likewise, excluding fully blocked configurations does not rule out the existence of distinct nontrivial communicating classes. To address this point, we tested the SAMC dynamics on exhaustively enumerable systems, where the full admissible configuration space is known exactly.

Specifically, we considered: (i)  $3\times 3$ systems with $n=1$, $2$, $3$, $4$ linear chains, and (ii) $4\times 4$ systems with $n=1$, $7$ linear chains. For the  $3\times 3$ systems, the number of states was equal to  $M=1206$, $2592$, $1350$, $144$ for $n=1$, $2$, $3$, $4$, respectively. For the  $4\times 4$ systems, the number of states were equal to  $M=98784$, $15936$ for $n=1$, $7$, respectively. For the $3\times 3$ systems, Supplementary Figure \ref{fig:fig_SI_ergodicity}a---d show the state-by-state population histograms constructed from $10^6$ independent configurations collected in a single SAMC run for each value of $n$. The horizontal red line marks the expected mean population for uniform sampling, and the orange band indicates the corresponding standard deviation. The agreement of the observed statistics with the expected one in these testable cases supports the uniformity and ergodicity of SAMC. Note that the consistency between the observed and expected deviations from the average confirms a posteriori the independence of the samples. The latter were collected with a conservative time lag of $3 \tau_{eq}$, for which the edge-state overlap correlation function has decayed to a negligible value, see Supplementary Note 5.

Because $4\times 4$ systems have a much larger number of states than $3 \times 3$ ones, it would be impractical to assess SAMC ergodicity and uniformity using state-by-state histograms. Accordingly, we instead adopted the uniformity and ergodicity test introduced in Ref.~\cite{mansfield2006unbiased}. The test consists of computing the quantity $m_k$, defined as the number of distinct conformations that appear exactly $k$ times in $K$ statistically independent samples, and to compare it with the expected binomial distribution for unbiased uniform sampling of the $M$ possible states. Indicating with $p=1/M$ the uniform sampling probability of the states we have
\begin{equation}
m_k^{\rm theor}=M\, p^k \, (1-p)^{K-k}\, K!/(k! (K-k)!).
\label{eq:mansfield}
\end{equation}

The data points in panels (e-f) of Supplementary Figure \ref{fig:fig_SI_ergodicity} correspond to the observed $m_k$ curves obtained by collecting $K=10^8$ independent samples of $4 \times 4$ systems with $n=1$ ($M=98784$) and $n=7$ ($M=15936$), using a single SAMC trajectory in each case. The solid curves indicate the expected $m_k^{\rm theor}$ based on the theoretical expressions of eq.~\ref{eq:mansfield}. The observed and expected curves are fully consistent.

Taken together, these results provide numerical evidence that, for the exhaustively testable systems considered here, (i) SAMC sampling is uniform and (ii) the admissible configuration space does not break into disconnected components under the SAMC dynamics, supporting the effective ergodicity and uniformity of SAMC in the considered regimes.

\subsection{Rejection rate.}
\begin{figure}[t!]
\centering
\includegraphics[width=0.41\textwidth]{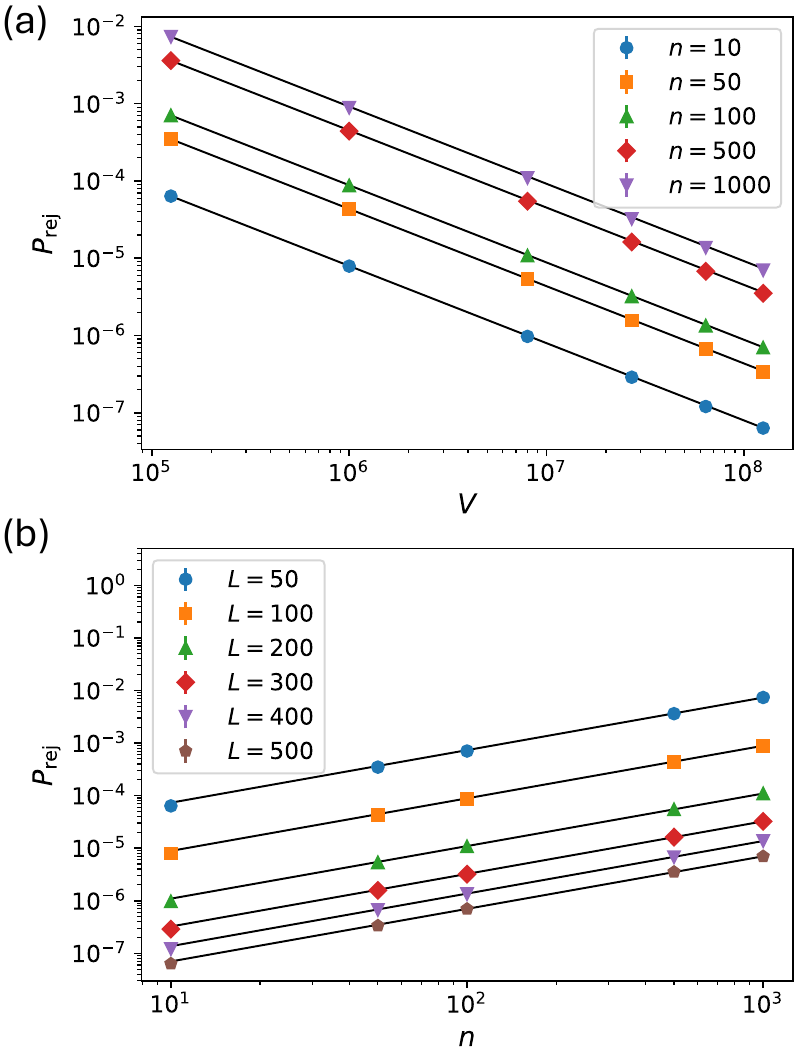}
\caption{{\bf Statistics of SAMC move rejections.} Probability of rejection of SAMC moves for various combinations of system sizes $V=L^3$ and number of linear chains $n$. Panel (a) shows $P_{\rm rej}$ as a function of $V$ for different values of $n$. The solid line is a best fit to the data using an inverse proportionality of $P_{\rm rej}$ on $V$. Panel (b) shows $P_{\rm rej}$ as a function of $n$ for different lengths of the box side $L$. The solid line is a best fit to the data using a direct proportionality of $P_{\rm rej}$ on $n$. The data and the best fits are consistent with the rejection probability $P_{\rm rej}$ scaling asymptotically as $O(n/V)$, the expected probability that site $C$ is an endpoint. $P_{\rm rej}$ data points were obtained from the block analysis (10 sets) of SAMC trajectories of $10^7 - 10^8$ Monte Carlo steps. Error bars represent standard errors.}
\label{fig:fig_SI_rej}
\end{figure}

Supplementary Figure \ref{fig:fig_SI_rej} shows the rejection probability $P_{\rm rej}$ of elementary SAMC moves for various combinations of system size $V=L^3$ and number of linear chains $n$. The data and the best fits are consistent with the rejection probability $P_{\rm rej}$ scaling asymptotically as $O(n/V)$, the expected probability that site $C$ is an endpoint.

\vfill
\eject
\subsection{Decorrelation}
\label{sec:decorrelation}
\begin{figure}[t!]
\centering
\includegraphics[width=0.45\textwidth]{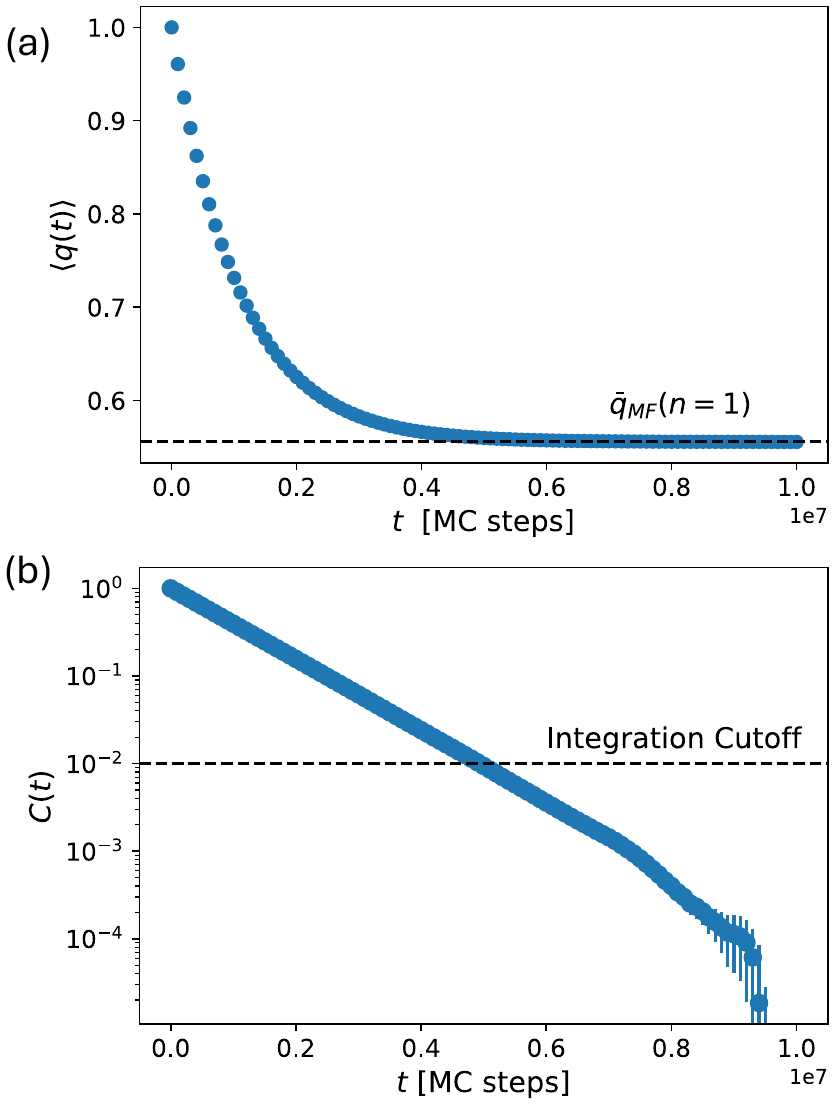}
\caption{{\bf Autocorrelation of the edge–state overlap.} (a) Average edge-state overlap $\langle q(t) \rangle$ between two configurations separated by $t$ Monte Carlo steps. The asymptotic value is consistent with the mean-field estimate $\bar{q}_{\rm MF}(n=1) \simeq 5/9$ (dashed line). The semi-log plot in panel (b) shows the corresponding normalized autocorrelation function $C(t)$. The dashed line corresponds to the $10^{-2}$ threshold defining the cut off of the $C(t)$ integral used to compute the decorrelation time, $\tau_{\rm eq}$. Data in panels (a-b) are for a system of size $100 \times 100 \times 100$ and $n = 1$, and are obtained by averaging over $m=100$ pairs of configurations for each $t$. At large $t$ the statistical uncertainty of the blocked average of $q$, and hence $C(t)$, is about $10^{-5}$, consistent with the observed onset of the noise in the tail of $C(t)$. Mean-field arguments (see text) for the statistical error on $q$ yield an order of magnitude estimate of $(m \cdot 3V)^{-1/2} \sim 10^{-5}$, comparable to the observed one. After integration, the relative uncertainty of $C(t)$ is inherited by $\tau_{\rm eq}$. Error bars represent standard errors.}
\label{fig:fig_SI_autocorrelation}
\end{figure}

Supplementary Figure \ref{fig:fig_SI_autocorrelation}a shows the average edge-state overlap $\langle q(t) \rangle$ between two configurations separated by $t$ Monte Carlo steps for $100 \times 100 \times 100$ systems with $n = 1$. \\
Asymptotically, $\langle q(t) \rangle$ approaches a plateau value $\bar{q}$, which can be estimated with the following theoretical mean-field argument. Each configuration has $N_{\rm occupied} = V - n$ edges occupied by bonds and $N_{\rm empty} = d\,V - N_{\rm occupied}$ unoccupied edges, where $d$ is the lattice dimensionality. Let $x_n = N_{\rm occupied}/(d\,V) = (1 - n/V)/d$ be the fraction of occupied edges, then the mean fraction of edges that have the same state (either occupied or empty) in two independent configurations is $\bar{q}_{\rm MF}(n) = 1 - 2x_n(1-x_n)$.\\
In our case, where $d=3$ and $n \ll V$, the expression reduces to $\bar{q}_{\rm MF}(n) = 5/9 + O(n/V)$, which approximates well the value of $\bar{q}$ calculated from the data, see Supplementary Figure \ref{fig:fig_SI_autocorrelation}a.\\
The same mean-field argument can be used to estimate the statistical uncertainty of numerical evaluations of the edge-state overlap of two uncorrelated configurations. Considering $q$ as the sum of $dV$ independent and identically-distributed bond-overlaps and that it is normalized to $d\,V$, one has that the estimated uncertainty of $q$ scales as $(d\,V)^{-1/2}$.

The normalized autocorrelation of the edge-state overlap, $C(t) = (\langle q(t) \rangle -\bar{q}) / (1 - \bar{q})$, is shown in Supplementary Figure \ref{fig:fig_SI_autocorrelation}b. Its linearity in the semi-log plot, modulo finite-size effects, indicates that it is well approximated by a single exponential decay. The decorrelation (decay) time scale,  $\tau_{\rm eq}$, was estimated by integrating $C(t)$ up to the smallest time lag $t$ for which $C(t) < 10^{-2}$.

\begin{figure}[t!]
\centering
\includegraphics[width=0.45\textwidth]{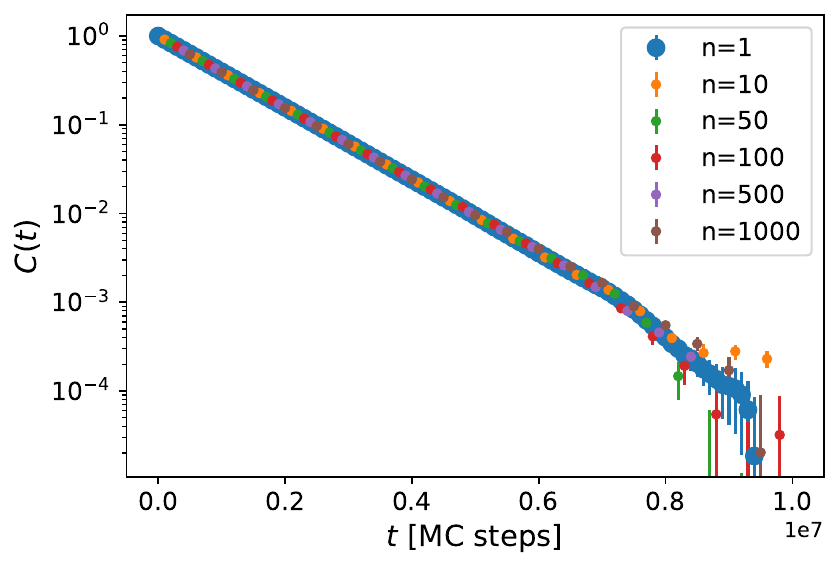}
\caption{{\bf Decorrelation time and number of linear chains.} Semi-log plot of the normalized autocorrelation function $C(t)$ for a system of size $100 \times 100 \times 100$ and various numbers of linear chains $n$. Each data point was obtained by averaging over 100 pairs of configurations. Despite the broad range of $n$, the data points fall on the same $C(t)$ curve, indicating the lack of a significant dependence of the decorrelation (decay) time $\tau_{\rm eq}$ on $n$ at fixed $V$. Error bars represent standard errors.}
\label{fig:fig_SI_taumc_vs_n}
\end{figure}

Supplementary Figure \ref{fig:fig_SI_taumc_vs_n} shows that the normalized autocorrelation function $C(t)$ for $100\times 100\times 100$ systems does not change appreciably as $n$ varies from $1$ to $1000$. Accordingly, $\tau_{\rm eq}$, obtained by integrating $C(t)$, is also practically independent of $n$.

Finally, in Supplementary Figure \ref{fig:fig_SI_taueq} we present the single-core CPU time per elementary SAMC step, showing that it has only a weak dependence on system size, until memory requirements exceed the (machine-dependent) RAM capacity, at which point memory handling overheads set in.
\vfill

\vfill
\pagebreak
\eject

\begin{figure}[t!]
\centering
\includegraphics[width=0.45\textwidth]{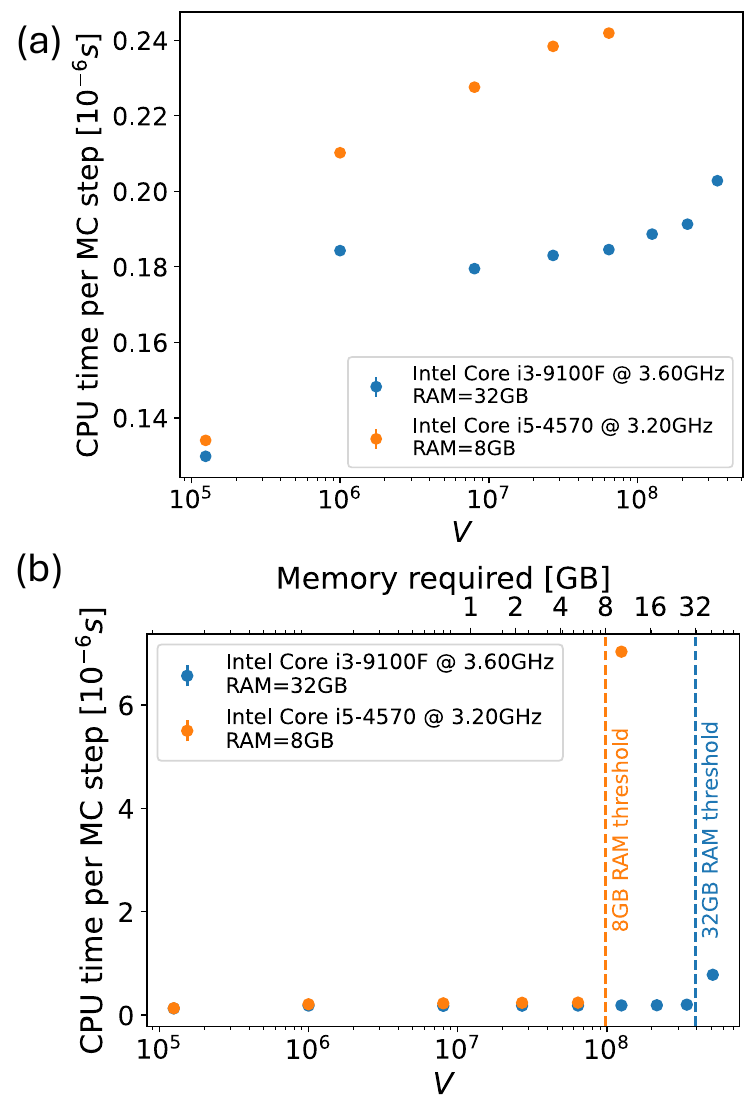}
\caption{{\bf CPU time per Monte Carlo step.} Single-core CPU time per elementary Monte Carlo step as a function of system size $V$ and $n=1$, measured on two standard workstations with Unix operating systems with different processors and RAM quantity, as indicated. (a) The CPU time per SAMC step is approximately independent of $V$ until the memory requirements exceed the RAM capacity of the machine, at which point memory-handling overheads significantly degrade performance. The performance dependence on $V$ when RAM capacity is not exceeded is illustrated in more detail in panel (b), which shows that the CPU time per SAMC step has a weak dependence on $V$, increasing by only about a factor of $2$ over a $1000$-fold increase of $V$. For each system size $V$, the CPU time per Monte Carlo step was estimated from SAMC runs performed without writing configurations to disk. The run durations spanned increasing fractions of $\tau_{\rm eq}$, from $0.05\tau_{\rm eq}$ to $0.25\tau_{\rm eq}$, and the resulting timings were linearly interpolated to obtain a reliable estimate of the per-step CPU cost while discounting initialization overheads, including memory allocation. Error bars, consistently smaller than the symbols, represent standard errors.}
\label{fig:fig_SI_taueq}
\end{figure}

\subsection{SAMC variants}
\subsubsection{Lower-rejection SAMC variant}
The efficiency of the self-assembly Monte Carlo can be enhanced with the following change of the elementary move.
\begin{itemize}
\item[(1)] {Endpoint selection.} Choose uniformly one of the $2n$ endpoints, denoted as $A$.
\item[($2^\prime$)]{Neighbour selection.}
Choose uniformly a non-bonded nearest neighbour of $A$, denoted $B$. Next, choose uniformly one of the bonded neighbours of $B$ that is not an endpoint, denoted $C$. If no such site exists, reject the move, retain the current configuration.
\item[(3)] {Bond swap.} Replace bond $BC$ with $AB$.
\end{itemize}
\noindent In this variant, the choice of site $C$ is conditioned on the bonded neighbours of $B$ that are not endpoints, if any.

{\bf Detailed balance and uniform stationary measure.} The conditioned selection at step $(2^\prime)$ preserves detailed balance with respect to the uniform stationary measure on each connected component.
Let us consider the forward and reverse move between admissible configurations, $\Gamma$ and $\Gamma^\prime$, differing by the $BC \leftrightarrow AB$ bond swaps.

Based on considerations analogous to those presented Supplementary Note 1, the forward and reverse proposal probabilities are
\begin{eqnarray}
q(\Gamma\to\Gamma^{\prime})&=\frac{1}{2n}\,\frac{1}{2d-1}\,\frac{1}{\tilde{m}_\Gamma(B)},\\
q(\Gamma^\prime\to\Gamma)&=\frac{1}{2n}\,\frac{1}{2d-1}\,\frac{1}{\tilde{m}_{\Gamma^\prime}(B)},
\end{eqnarray}
\noindent where $\tilde{m}_\Gamma(B)$ denotes the number of bonded neighbours of $B$ that are not endpoints.
Next, we note that with the $BC\leftrightarrow AB$ bond swaps, the bonded neighbourhood of $B$ changes solely for sites $A$ and $C$ being traded as bonded neighbours of $B$. Since $C$ is not an endpoint in $\Gamma$ and $A$ is not an endpoint in $\Gamma^{\prime}$, the number of bonded neighbours of $B$ that are not endpoints is unchanged,
\begin{equation}
\tilde{m}_\Gamma(B)=\tilde{m}_{\Gamma^{\prime}}(B).
\end{equation}

We thus have that the forward and reverse proposal probabilities are the same
\begin{equation}
q(\Gamma\to\Gamma^{\prime})=q(\Gamma^{\prime}\to\Gamma).
\end{equation}
Because all admissible proposals are accepted, and rejected proposals again contribute only to self-loops, the conditioned variant also satisfies detailed balance with respect to the uniform measure on each connected component.

{\bf Efficiency enhancement.}
The rejection rate of this variant is much lower than in the formulation of Supplementary Note 1 because rejection can now occur only if: (i) $B$ is an endpoint of a chain of length 1, so that its unique bonded neighbour is an endpoint, or (ii) $B$ is the midpoint of a chain of length 2, so that both its bonded neighbours are endpoints. Thus, rejections require two distinct endpoints in the neighbourhood of $A$. The rejection rate thus scales as $O(n^2/V^2)$, hence lower than the original formulation.

\subsubsection{SAMC with plaquette-flip moves}

The native move set of the self-assembly Monte Carlo can be expanded with plaquette-flip moves.
The additional moves consist of:
\begin{itemize}
\item[(1)] randomly picking one plaquette;
\item[(2)] consider the two pairs of parallel edges of the plaquette. If one pair is occupied (two parallel bonds) and the other is empty (no bonds), swap the occupied and empty states of the edges. Otherwise reject the move.
\end{itemize}
An example is shown in Supplementary Figure \ref{fig:fig_SI_flip}.

The move set has been previously used to evolve systems exclusively consisting of rings on fully-packed periodic hypercubic lattices\cite{mansfield2006unbiased,slongo2023quantum}, for which it has been argued to be ergodic\cite{mansfield2006unbiased}, a property that may not hold with general boundary conditions, see Supplementary Section 9 of Ref.~\cite{slongo2023quantum}.

Alternating the original SAMC move set with plaquette-flip moves can evolve the system even away from the endpoints.

\begin{figure}[t!]
\centering
\includegraphics[width=0.45\textwidth]{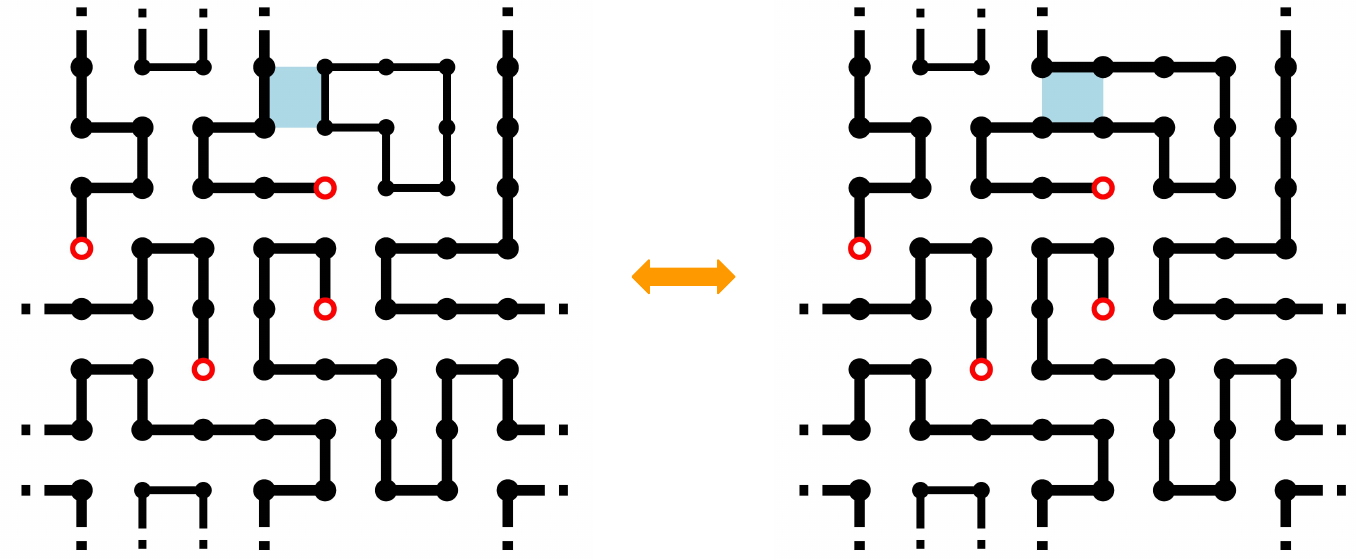}\\
\caption{{\bf Plaquette-flip moves.} In a plaquette-flip move, a randomly picked plaquette in which two parallel edges are occupied and the other two are empty (highlighted) is rotated by 90$^\circ$, so to flip the occupied/empty states of the edges.}
\label{fig:fig_SI_flip}
\end{figure}

\subsection{Giant linear chains}
\begin{figure}[t!]
\centering
\includegraphics[width=0.45\textwidth]{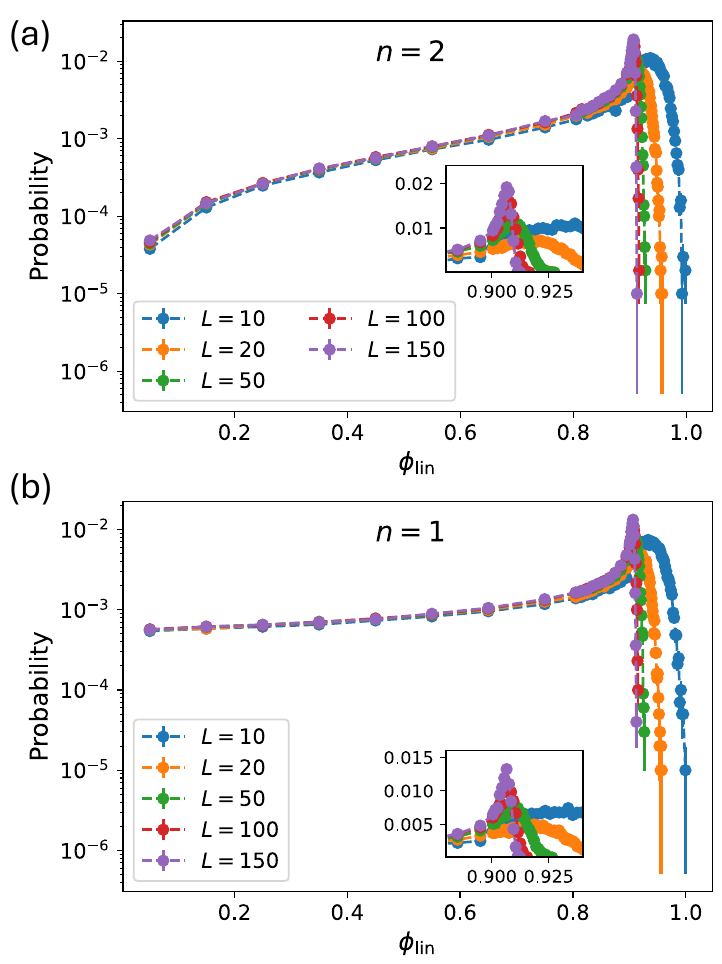}\\
\caption{{\bf Statistics of the volume fraction occupied by linear chains.} Probability distribution of the linear volume fraction $\phi_{\rm lin}$ for various system sizes $V$ with (a) $n = 2$ and (b) $n=1$ linear chains. $\phi_{\rm lin}$ approaches a limiting distribution for increasing system size, with $100\times100\times100$ systems being already effectively asymptotic. Each distribution was computed from $10^5$ independent configurations. Error bars represent standard errors.}
\label{fig:fig_SI_philin_distribution}
\end{figure}

\begin{table}[H]
\centering
\begin{tabular}{|l|l|}
\hline
\textbf{$\;n\;$} &\phantom{$X^{X^X}$} \textbf{$\rho(\bar{\phi}_{\rm lin})$ \phantom{$X^{X^X}$}} \\ \hline
$\;1\;$ & $\;\;0.0110 \pm 0.0003$ \\ \hline
$\;2\;$ & $\;\;0.0160 \pm 0.0004$ \\ \hline
$\;5\;$ & $\;\;0.0257 \pm 0.0005$ \\ \hline
$\;10\;$ & $\;\;0.0369 \pm 0.0003$ \\ \hline
$\;20\;$ & $\;\;0.0495 \pm 0.0007$ \\ \hline
\end{tabular}
\caption{{\bf Scaling of the modal volume fraction of linear chains.} Fraction of configurations, $\rho$, for which the linear volume density, $\phi_{\rm lin}$, falls in the $[0.90,0.91]$ interval. This selection serves as a practical and stringent criterion for conditioning systems at the modal linear volume fraction, $\bar{\phi}_{\rm lin}=0.907 \pm 0.001$. Data are for $100 \times 100 \times 100$ systems with varying number of linear chains $n$, as indicated.
$\rho(\bar{\phi}_{\rm lin})$ grows approximately as $\sqrt{n}$, consistent with the fact that the width of the  probability distribution of the linear chain fractions (Dirichlet partitioning) scales as $n^{-1/2}$. Indicated uncertainties represent standard errors.}
\label{tab:modal_linear_density}
\end{table}
\begin{figure}[t!]
\centering
\includegraphics[width=0.45\textwidth]{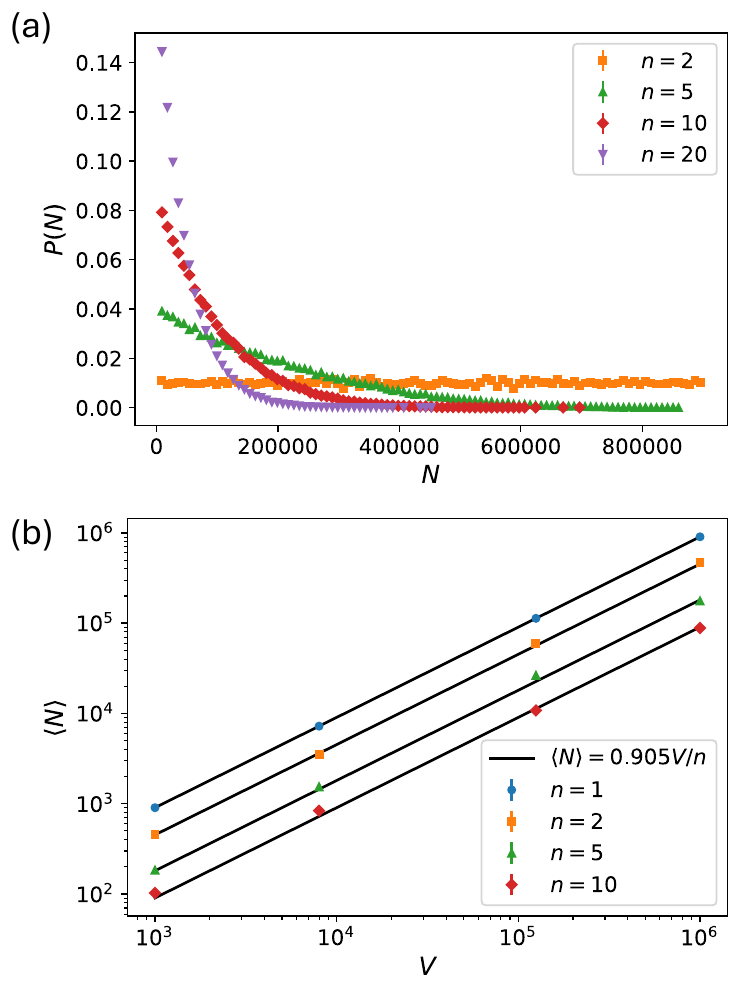}
\caption{{\bf Linear chain lengths at $\bar{\phi}_{\rm lin}$.} (a) Probability distribution of the length of individual giant chains for $100 \times 100 \times 100$ systems conditioned at $\bar{\phi}_{\rm lin}$ and with $n=2$, 5, 10, and 20. The plot is a different representation of the same data of Figure 3b in the main text. Data are from $10^5$ independent configurations for each $n$. (b) Mean individual chain length, $\langle N \rangle$ as a function of system size $V$ for different values of $n$, as indicated. The data fall on the solid lines which correspond to the parameter-free theoretical expressions for uniform equipartition of the linear volume among the $n$ chains, $\langle N \rangle^{\rm theor} = V_{\rm lin}/n=\bar{\phi}_{\rm lin} V/n$, with $\bar{\phi}_{\rm lin}$ set to 0.905, the midpoint of the conditioning interval $[0.90,0.91]$. Data are from $10^2$ independent configurations for each $(V,n)$ combination. Error bars, consistently smaller than the symbols, represent standard errors.}
\label{fig:fig_SI_chain_length}
\end{figure}

\begin{figure}[t!]
\centering
\includegraphics[width=0.45\textwidth]{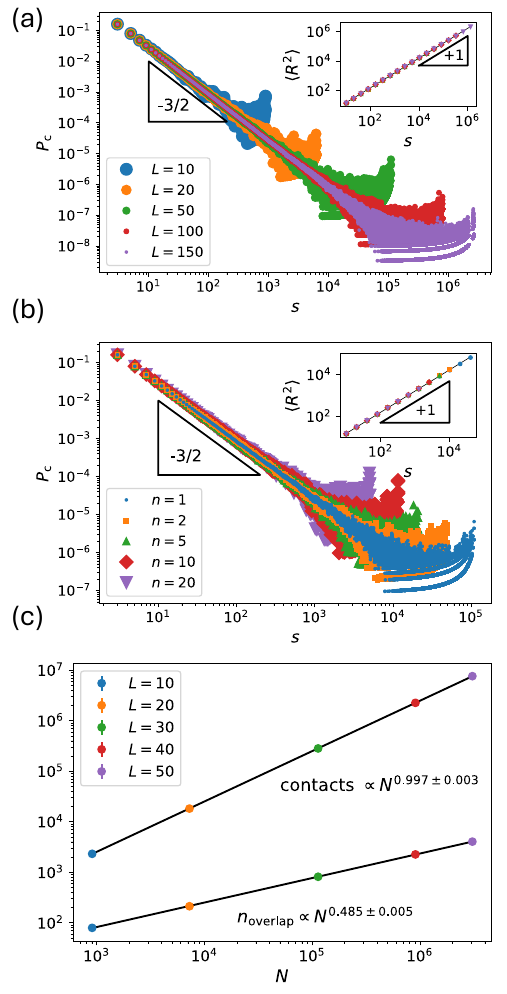}
\caption{{\bf Statistics of giant linear chains.} (a,b) Contact probability $P_c$ and mean square internal distance $\langle R^2 \rangle$ (inset) versus contour separation $s$ of giant chains for (a) systems of various sizes as indicated and with $n=1$ and (b) systems of size $50\times 50 \times 50$ and various $n$, as indicated.(c) Average total number of inter-chain contacts and average number of distinct overlapping chains as a function of chain length, $N$, for different system sizes as indicated and $n=1$. The indicated exponents of power-law best fits (solid lines) are consistent with melt behaviour. In all panels, data are from $10^2$ independent configurations conditioned at $\bar{\phi}_{\rm lin}$ and with near monodisperse chains. To enforce the latter constraint for $n>1$, we retained from the conditioned configurations only those individual giant chains with lengths within 10\% of the nominal mean value $\langle N \rangle^{\rm theor} = \bar{\phi}_{\rm lin} V/n$. Error bars in panel (c) represent standard errors.}
\label{fig:fig_SI_metric}
\end{figure}

\vfill
\eject
\subsection{Physical knots}
As detailed in the main text, the physical knotted state of the giant linear chains was established by computing topological invariants on circularized versions of the linear chains, obtained by applying the minimally invasive closure procedure of Ref.~\cite{tubiana2018kymoknot}.
Details of the resulting knot spectrum are given in Supplementary Table \ref{tab:spectrum}.

\begin{table}[t!]
    \centering
\begin{tabular}{|c|r|r|r|}
\hline
\hline
      {Knot type} &$L=10$& $L=15$ & $L=20$ \\
      \hline
       $0_1$ & 8891 & 5925  & 2496 \\
       $3_1$ & 866 & 2314 & 2361 \\
       $4_1$ & 109 & 341 & 391 \\
       $5_1$ & 31 & 85 & 105\\
       $5_2$ & 33 & 190 & 219 \\
       $6_1$ & 8 & 45 & 56 \\
       $6_2$ & 12 & 42 & 56 \\
       $6_3$ & 4 & 25 & 29 \\
       Other primes & 13 & 167 & 279 \\
       $3_1\#3_1$ & 22 & 415 & 1195 \\
       $3_1\#4_1$ & 6 & 128 & 360\\
       $3_1\#3_1\#3_1$ & 1 & 36 & 375 \\
       {\small Other composites}& 4 & 285 & 2024 \\
       Undecided & 0 & 2 & 53 \\
       \hline
       \hline
       {Knot type} & $L=25$ & $L=30$ & \\
       \hline
       $0_1$ & 584 & 62 & \\
       $3_1$ & 1114 & 214 & \\
       $4_1$ & 219 & 45 & \\
       $5_1$ & 43 & 15 & \\
       $5_2$ & 93 & 24 & \\
       $6_1$ & 28 & 1 & \\
       $6_2$ & 18 & 9 & \\
       $6_3$ & 24 & 1 & \\
       Other primes & 103  & 739 &  \\
       $3_1\#3_1$ & 1087 & 342 & \\
       $3_1\#4_1$ & 396 & 109 & \\
       $3_1\#3_1\#3_1$ & 666 & 396 & \\
       {\small Other composites}& 5403 & 7723 &\\
       Undecided & 222 & 322 & \\
       \hline
    \end{tabular}
    \caption{{\bf Knot spectrum of giant chains in $L \times L \times L$ systems of indicated size, conditioned at $\bar{\phi}_{\rm lin}$, and with $n=1$.} The data are the counts of the simplest knot types over sets of $10^4$ independent configurations. More complex knot types are gathered under the "other" label, except for the few instances, labeled as "undecided" for which the topology could not be assigned due to the complexity of the pattern of projected crossings after the geometrical simplifications.}
    \label{tab:spectrum}
\end{table}

{\bf Characteristic chain length for trefoil occurrence.} We obtained the characteristic chain length for trefoil occurrence, $\bar{N}_{\rm tref}$ (not to be confused with the trefoil knot length, $l_{\rm tref}$) by fitting the probability of trefoil knots, $P_{\rm tref}(N)$, with the theoretical expression of Ref.~\cite{deguchi1994statistical}, $P_{\rm tref}^{\rm theor}(N) = C_{\rm tref} \cdot (N/\bar{N}_{\rm tref})\cdot \exp{(-N/\bar{N}_{\rm tref})}$. As shown in Supplementary Figure \ref{fig:fig_SI_Ptref}, the best-fit yielded characteristic   $\bar{N}_{\rm tref} \simeq 4998 \pm 192$.  Notice that the $\bar{N}_{\rm tref}$ also coincides with the peak location of $P_{\rm tref}^{\rm theor}(N)$.

\begin{figure}[t!]
\centering
\includegraphics[width=0.45\textwidth]{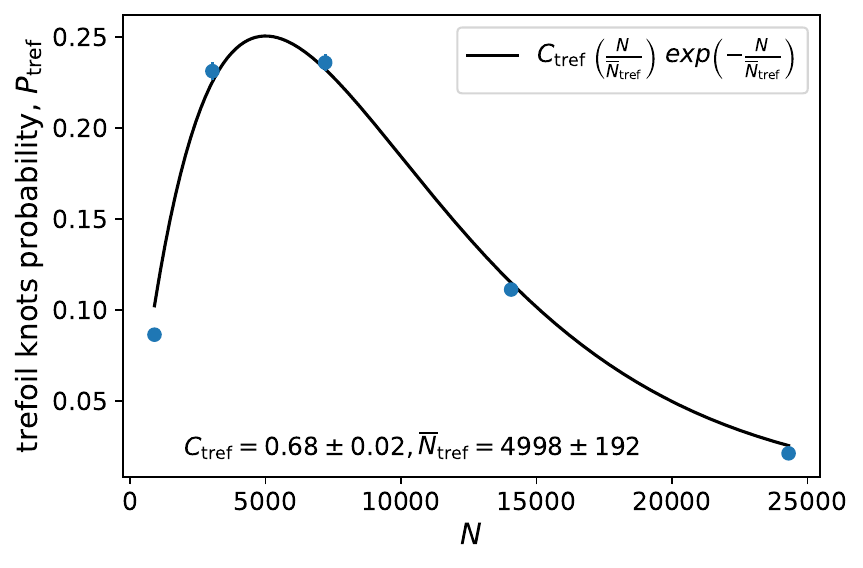}
\caption{{\bf Characteristic chain length for trefoil occurrence.} Probability of trefoil knots $P_{\rm tref}$ in giant chains as a function of their length $N$, based on Supplementary Table \ref{tab:spectrum}. The solid line is the best fit to the data using the theoretical expression of Ref.~\cite{deguchi1994statistical}, $P_{\rm tref}^{\rm theor}(N) = C_{\rm tref} \cdot (N/\bar{N}_{\rm tref})\cdot \exp(-N/\bar{N}_{\rm tref})$. The best-fit characteristic length is $\bar{N}_{\rm tref} \simeq 4998 \pm 192$.}
\label{fig:fig_SI_Ptref}
\end{figure}

\vskip 0.5cm

{\bf Search procedure for individual trefoil-knotted components.} The characteristic length of individual trefoil-knotted components in giant linear chains is determined with the following bottom-up search procedure.

\begin{enumerate}
    \item For each considered chain length $N$, we first restrict the analysis to giant linear chains whose closed versions feature exclusively one or more trefoil knots, i.e.~knots of type $3_1$, $3_1\#3_1$, $3_1\#3_1\#3_1$, etc.

    \item Starting from the chain end with monomer index 1, we move towards the chain interior to identify the first locally exposed monomer, denoted $i$. By ``locally exposed'' we mean that the monomer is linearly separable from the following $\bar{s}=200$ monomers, or fewer if the opposite chain end is closer. The linear separability is tested with a perceptron algorithm. It serves to establish whether there exists at least one plane passing through $i$ such that all other monomers lie on the same side of the plane.

    \item Starting from $i$, we then consider chain segments of increasing length by moving the second endpoint in steps of $\Delta S=25$ monomers towards the chain interior. At each extension step, we test whether the terminal monomer $j$ of the interval is locally exposed, i.e.~linearly separable from the preceding $\bar{s}=200$ monomers, or fewer if $i$ is closer.

    \item If $j$ is not locally exposed, we enlarge the segment by a further $\Delta S$ monomers and repeat the test. If $j$ is locally exposed, we apply the minimally invasive closure scheme to the interval bracketed by the exposed ends $i$ and $j$, and test whether the resulting physical knotted state is a trefoil knot. Requiring local exposure at both termini provides robustness to the closure procedure by ensuring that each terminus admits a direction of extension that does not interfere with its surrounding neighborhood.

    \item If the segment is not trefoil-knotted, we continue the segment-extension procedure. If the test is successful, we further shorten the knotted segment by moving each terminus, in turn, towards the segment interior in steps of $\delta S=5$ monomers. At each reduction step, we test whether the reduced termini remain locally exposed and, if so, whether the trefoil knot persists after closure. The shortest interval still yielding a trefoil knot is taken as the corresponding trefoil-knotted component.

    \item The same search procedure is then reapplied to the remainder of the chain, using monomer $j+1$ as the new starting point. The search terminates when $j$ reaches the opposite chain end.

    \item Finally, we compare the number of identified trefoil-knotted components with the number of trefoil components found for the entire closed chain. Only matching cases are retained for computing the contour lengths of the trefoil-knotted components. Depending on $N$, the retained fraction ranged from about 95\% to 50\%.
\end{enumerate}

\subsection{Gauss linking integral}
The Gauss linking integral between two self-avoiding, non-crossing curves $\gamma_a(t_a)$ and $\gamma_b(t_b)$ in three dimensions is defined as
\begin{align}
    &L_G[\gamma_a, \gamma_b] =\\
    &= \nonumber
    \frac{1}{4 \pi} \int_0^1 dt_a \int_0^1 dt_b \;
    \frac{( \dot{\gamma_a}(t_a) \times \dot{\gamma_b}(t_b)) \cdot (\gamma_a(t_a) - \gamma_b(t_b))}{|\gamma_a(t_a) - \gamma_b(t_b)|^3}\ .
\end{align}
This integral admits a closed-form expression when both curves are piecewise-linear. Consider first the case when both curves are straight segments, $s_a$ and $s_b$, with coordinates parametrized as
\[
\mathbf{s}_k(t) = \mathbf{r}_{k1} + t\,(\mathbf{r}_{k2} - \mathbf{r}_{k1}), \qquad k = a,b\ .
\]
In this case, the evaluation of the Gauss linking integral $L_G[s_a, s_b]$ yields a sum of four oriented solid angles:
\begin{equation}
    L_G[s_a, s_b] = \frac{1}{2 \pi} (-\theta_{11} + \theta_{12} + \theta_{21} - \theta_{22})
\end{equation}
\noindent where
\begin{equation}
    \theta_{ij} = \arctan \frac{\mathbf{R}_{ij} \cdot (\hat{\mathbf{v}}_a \times \hat{\mathbf{v}}_b)}{|\mathbf{R}_{ij}| + |\mathbf{R}_{ij}| \hat{\mathbf{v}}_a \cdot \hat{\mathbf{v}}_b + \mathbf{R}_{ij} \cdot \hat{\mathbf{v}}_a + \mathbf{R}_{ij} \cdot \hat{\mathbf{v}}_b}\ ,
\end{equation}
and $\hat{\mathbf{v}}_k = (\mathbf{r}_{k2} - \mathbf{r}_{k1}) / |\mathbf{r}_{k2} - \mathbf{r}_{k1}| $, $\mathbf{R}_{ij} = \mathbf{r}_{ai} - \mathbf{r}_{bj}$, $i,j=1,2$.
\\
\\
With this expression, the evaluation of $L_G$ for two piecewise-linear chains is performed by summing the Gauss linking integral over all distinct pairs of segments, one from each chain.

The calculation of $L_G$ scales like the product of the two chains' lengths, and becomes impractical when each of the chains is composed of more than $10^5$ segments, as in our case. The numerical evaluation of $L_G$ can be significantly sped up by applying "topology-preserving" geometric simplifications to the chains, which can significantly reduce their lengths without introducing inter-chain crossings, thereby leaving $L_G$ mostly unchanged. To this end, we first iteratively applied on-lattice BFACF-type simplifications to both chains until no further reductions in chain length were possible. This step shortens both chains to about one-third of their original lengths, without introducing intra- and inter-chain crossings, while keeping all bonds on the lattice. Next, conservative off-lattice rectification moves were applied only to one of the two chains. In this second step, an entire sequence of consecutive bonds along one Cartesian direction, followed by one bond along a different axis, is replaced by a single "diagonal" (off-lattice) bond joining the endpoints.
The overall simplification provided a speed-up factor of about $15$, while leaving $L_G$ unchanged within $1$\%, as shown in Supplementary Figure \ref{fig:fig_SI_geom_simpl}.

The distribution of $L_G$ is shown in Supplementary Figure \ref{fig:fig_SI_linking_distribution}. The zero-inflated character of the histogram indicates that interlocking between any given pair of proximal chains is modest. In addition, the tails decay more slowly than a Gaussian, which would have a parabolic profile in the semi-log plot.
\vfill
\eject

\begin{figure}[t!]
\centering
\includegraphics[width=0.4\textwidth]{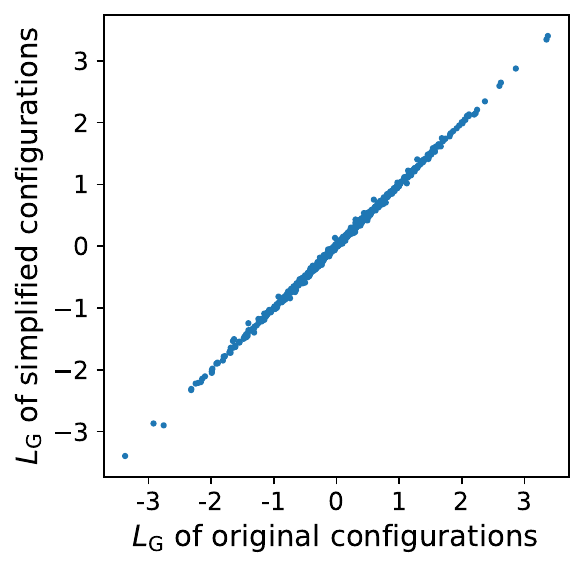}
\caption{{\bf Effect of the chain simplification on the Gauss linking integral $L_G$.} The scatter plot compares the Gauss linking integral $L_G$ for the minimal periodic images of two distinct giant chains without (horizontal axis) and with (vertical axis) geometric simplification of their backbones. Data are for $10^3$ independent configurations of a $10 \times 10 \times 10$ system conditioned at $\bar{\phi}_{\rm lin}$ and with $n = 2$. The data points lie close to the diagonal, with a relative mean-square difference smaller than $1\%$. Thus, the simplification does not significantly alter the value of $L_G$, while substantially speeding up its evaluation due to the shorter chains.}
\label{fig:fig_SI_geom_simpl}
\end{figure}

\begin{figure}[t!]
\centering
\includegraphics[width=0.45\textwidth]{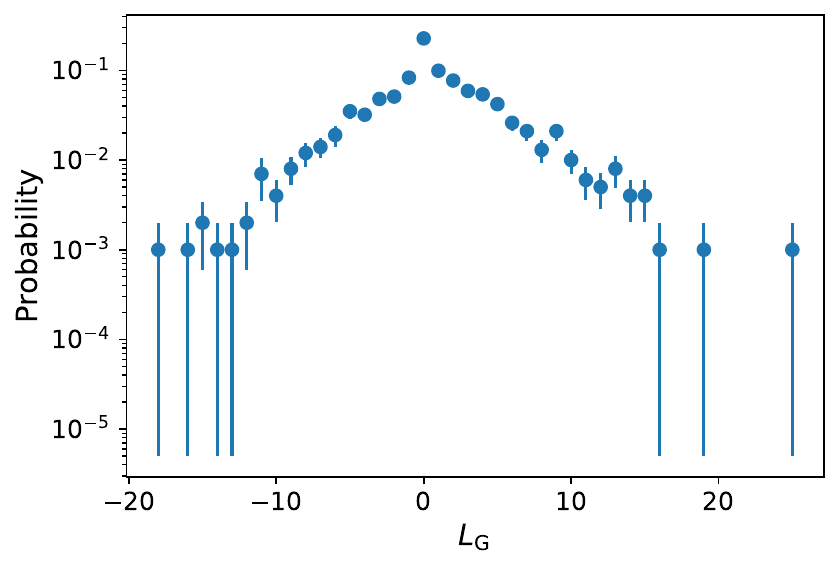}
\caption{{\bf Gauss linking integral distribution.} Histogram of the Gauss linking integral $L_G$ of giant chains in $100 \times 100 \times 100$ systems conditioned at $\bar{\phi}_{\rm lin}$ and with $n = 2$. $L_G$ is computed as the Gauss linking integral of the minimal periodic images of the two chains (the images with the closest centres of mass) over $10^3$ independent configurations. The reported $L_G$ is computed after chain simplification. For simplicity, data are binned at integer values, although $L_G$ for open curves is in general non-integer. Error bars represent standard errors.}
\label{fig:fig_SI_linking_distribution}
\end{figure}

\subsection{Ring background}
\begin{table}[t!]
\centering
\begin{tabular}{|l|l|}
\hline
\textbf{$\;n\;$} &\phantom{$X^{X^X}$} \textbf{$\langle l_{\rm ring} \rangle$ \phantom{$X^{X^X}$}} \\ \hline
$\;1\;$ & $\;\;\;\;\;8.84 \pm 0.08$ \\ \hline
$\;2\;$ & $\;\;\;\;\;8.91 \pm 0.04$ \\ \hline
$\;5\;$ & $\;\;\;\;\;8.79 \pm 0.02$ \\ \hline
$\;10\;$ & $\;\;\;\;\;8.82 \pm 0.02$ \\ \hline
$\;20\;$ & $\;\;\;\;\;8.78 \pm 0.02$ \\ \hline
\end{tabular}
\caption{{\bf Mean ring length.} Average ring length, $\langle l_{\rm ring} \rangle$, computed for $10^2$ independent configurations of $100 \times 100 \times 100$ systems conditioned at $\bar{\phi}_{\rm lin}$ and with various $n$, as indicated. Reported uncertainties represent standard errors.}

\label{tab:mean_ring_length}
\end{table}

Figure 3c of the main text shows the collapse onto a $-5/2$ power law of $\langle n_k \rangle /V$, the average number density of rings of length $k$, for configurations conditioned at the modal linear volume fraction $\bar{\phi}_{\rm lin}$. This scaling is accounted for by a simple theoretical argument that is more naturally formulated by normalizing the average number density to the volume of the ring background, $p_k = \langle n_k \rangle/[(1- \bar{\phi}_{\rm lin}) V]$. Because we use the same interval $\phi_{\rm lin} \in [0.90,0.91]$ to condition configurations at $\bar{\phi}_{\rm lin}$, $p_k$ and $\langle n_k \rangle /V$ differ solely by a prefactor that does not affect the scaling properties.

The number density $p_k$ can be estimated as follows. The probability that the two endpoints of a linear chain of $k-1$ steps are in contact, hence yielding a ring of length $k$, scales as $(k-1)^{-d\nu}$, where $d$ is the spatial dimension and $\nu$ is the Flory exponent, which in our case are $d=3$ and $\nu = 1/2$. Additionally, for cyclic symmetry, a given ring of length $k$ can be obtained by $k$ different linear chain closures. Overall, this gives $p_k^{\rm theor}= {\mathcal N} k^{-1}(k-1)^{-d\nu}$, where $\mathcal{N}$ is a normalizing prefactor such that $\sum_k k \; p^{\rm theor}_k = 1$. For $k\gg 1$, $p_k$ decays as $k^{-d \nu -1}$, which in our case specializes to $p_k \propto k^{-5/2}$. Due to this rapid decay, ${\mathcal N}$ is well approximated using an unrestricted summation over all possible ring lengths $k=4, 6, 8, ...,$ etc.:
\begin{equation}
    {\mathcal N} = \Bigg[\sum_{k \ge 4 \; {\rm even}} (k-1)^{-d \nu}\Bigg]^{-1} = \frac{1}{(1-2^{-d\nu})\,\zeta(d\nu) - 1}
\end{equation}

\noindent where $\zeta(x)$ is the Riemann zeta function.
Thus one has:
\begin{equation}
p_k^{\rm theor} = \frac{1}{(1-2^{-d\nu})\,\zeta(d\nu) - 1} \cdot  \frac{1}{k(k-1)^{d \nu}}\ ,
\end{equation}
\noindent and hence
\begin{equation}
\frac{\langle n_k \rangle^{\rm theor}}{V} = \frac{1 - \bar{\phi}_{\rm lin}}{(1-2^{-d\nu})\,\zeta(d\nu) - 1} \cdot  \frac{1}{k(k-1)^{d \nu}}
\end{equation}
\noindent which evaluates numerically to
\begin{equation}
\frac{\langle n_k \rangle^{\rm theor}}{V} \simeq 0.1379 \cdot \frac{1}{k(k-1)^{3/2}}\ .
\end{equation}
\noindent Figure 3c of the main text shows that the above expression provides an accurate parameter-free interpolation of the data points. Based on it, the average ring length $\langle l_{\rm ring} \rangle$ can be expressed analytically:
\begin{align}
&\langle l^{\rm theor}_{\rm ring} \rangle
= \frac{1}{\sum_k p^{\rm theor}_k}\\
&= \dfrac{(1-2^{-d \nu})\zeta(d \nu) - 1}{\sum\limits_{m=0}^{\infty} (-1)^m ((1-2^{-1-d \nu-m})\zeta(1+d \nu+m) - 1)}\ . \nonumber
\end{align}
Note that the expression holds independent of $V$, which is assumed to be large enough to allow for carrying out the unrestricted sum over ring lengths. Numerical evaluation of the analytical expression for $d=3$ and $\nu=1/2$ yields $\langle l^{\rm theor}_{\rm ring} \rangle \simeq 8.26$. The theoretical prediction is a good approximation to the observed average ring length $\langle l_{\rm ring} \rangle \simeq 8.8$, see Supplementary Table \ref{tab:mean_ring_length}.

\end{document}